\documentclass[showpacs,
aps,superscriptaddress, 10pt,
prd,notitlepage,showkeys,
nofootinbib,floatfix,twocolumn]{revtex4-2}
\usepackage{bm}
\usepackage{amsfonts}
\usepackage{latexsym}
\usepackage[normalem]{ulem}
\usepackage{graphicx}
\usepackage{amsmath}
\usepackage{palatino}
\usepackage{mathpazo}
\usepackage{tikz}
\usepackage{tikz-feynman}
\usepackage{textcomp}
\usepackage{float}
\usepackage{booktabs}
\usepackage{dcolumn}
\usepackage{ragged2e}
\usepackage{hyperref}
\hypersetup{colorlinks=true,citecolor=red,linkcolor=blue,filecolor=magenta,urlcolor=blue}
\usepackage{subfigure}
\DeclareUnicodeCharacter{2212}{-}
\usepackage{orcidlink}
\usepackage{epsfig}
\usepackage[justification=raggedright]{caption}
\usepackage[toc]{appendix}
\usepackage{commath}
\usepackage{cancel}
\usepackage{csquotes}
\usepackage{placeins}
\usepackage{multirow}
\usepackage{parskip}
\usepackage{threeparttable}
\usepackage{tcolorbox}
\usepackage{pifont}
\usepackage{siunitx}
\usepackage{soul, color, xcolor}
\usepackage{enumitem}
\setlist[itemize]{leftmargin=1em, itemindent=0em}

\begin{document}

\raggedbottom
\newcommand{\BITS}{Department of Physics, Birla Institute of Technology and Science - Pilani, K. K. Birla Goa Campus, NH-17B, Zuarinagar, Sancoale, Goa- 403726, India}

\newcommand{\WUSTL}{Physics Department and McDonnell Center for the Space Sciences, Washington University in St. Louis; Missouri, 63130, USA}

\newcommand{\Tezpur}{Department of Physics, Tezpur University, Napaam, Tezpur, 784028, Assam, India}

\newcommand{\ccnu}{Institute of Astrophysics, Central China Normal University, Wuhan 430079, China}
\newcommand{\IUCAA}{Inter-University Centre for Astronomy and Astrophysics, PostBag 4, Ganeshkhind, Pune-411007, Maharashtra, India}

\title{Revealing dark matter's role in neutron stars anisotropy: A Bayesian approach using multimessenger observations}

\author{Xue-Zhi Liu~\orcidlink{0009-0008-3286-7254}}\email{xz\_liu@mails.ccnu.edu.cn}
\affiliation{\ccnu}
\author{Premachand Mahapatra~\orcidlink{0000-0002-3762-8147}}\email{p20210039@goa.bits-pilani.ac.in}
\affiliation{\BITS}

\author{Chun Huang~\orcidlink{0000-0001-6406-1003}}\email{chun.h@wustl.edu}
\affiliation{\WUSTL}

\author{Ayush~Hazarika~\orcidlink{0009-0004-5255-0730}}\email{ayush.hazarika4work@gmail.com}
\affiliation{\Tezpur}

\author{Chiranjeeb Singha~\orcidlink{0000-0003-0441-318X}}\email{chiranjeeb.singha@iucaa.in}
\affiliation{\IUCAA}

\author{Prasanta Kumar Das~\orcidlink{0000-0002-2520-7126}}\email{pdas@goa.bits-pilani.ac.in}
\affiliation{\BITS}

\date{\today}

\begin{abstract}
Dark matter (DM) continues to evade direct detection, but neutron stars (NSs) serve as natural laboratories where even a modest DM component can alter their structure. While many studies have examined DM effects on NSs, they often rely on specific choices of equations of state (EOS) models, assume isotropy, and lack a Bayesian statistical framework, limiting their predictive power.  In this work, we present a Bayesian framework that couples pressure-anisotropic nuclear EOS to a self-interacting fermionic DM component, constrained by NICER and GW170817 data. Our results show that DM mass fractions up to $\sim10\%$ remain consistent with current data, which softens the high-density EOS, leading to reduced stellar radii and tidal deformabilities while requiring negligible pressure anisotropy. Bayesian model comparison reveals no statistically significant preference between pure baryonic and DM-admixed NSs, indicating that DM inclusion enhances physical realism without complexity penalties. However, existing data cannot tightly constrain the DM parameters, and our empirical radius definition introduces a systematic bias toward the DM core configurations. To address this, we therefore introduce the DM radius span $\Delta R_\chi \equiv R_{\chi,\mathrm{max}} - R_{\chi,\mathrm{min}}$ as a unified diagnostic for DM distributions. This parameter simultaneously characterizes core-halo transition features while exhibiting strong linear correlations ($\Delta R_\chi < \SI{4}{km}$) with both DM and baryonic parameters, providing a clear avenue for future constraints. Our approach bridges current limitations and future potential in probing DM through compact star observations.
\end{abstract}

\maketitle

\section{Introduction}
Dark matter (DM) is an unseen form of matter that dominates the Universe's mass budget; in the concordance $\Lambda$ cold dark matter framework it makes up roughly $85\%$ of all matter, about $27\%$ of the total cosmic energy density while the remainder is baryonic matter (BM) and dark energy \cite{Bertone:2016nfn}. Despite decades of intensive experimental searches, no experiment has yet achieved a conclusive direct detection of DM particles. Recently, complementary astrophysical probes including electromagnetic observations \cite{DeRocco:2019njg, MADMAX:2024jnp, Firouzjahi:2020whk}, high-energy neutrinos \cite{Rembiasz:2018lok}, and gravitational-wave measurements \cite{Guo:2022dre, Miller:2025yyx, Marriott-Best:2025sez} have been proposed as a multi-messenger strategy for investigating DM physics.

Various hypothetical particles have been suggested as DM candidates, including feebly interacting massive particles, as well as neutrinos with very weak interaction strengths~\cite{Datta:2021elq}. Recently, there has been a surge of interest in low-mass or ultralight DM candidates, particularly axions and axionlike particles (ALPs)~\cite{Ringwald:2024uds}. These particles are theoretically well-motivated and arise naturally in various extensions of the Standard Model, including solutions to the strong \textit{CP} problem and string theory frameworks. Over the past years, a wide range of experimental and observational efforts have been dedicated to detecting such particles, from laboratory-based searches to astrophysical and cosmological probes. These efforts have significantly constrained the allowed parameter space for axions and ALPs, placing stringent bounds on their masses and interaction strengths~\cite{Schumann:2019eaa}. 
Among the many DM candidates, which are weakly interacting massive particles (WIMPs)~\cite{Cooley:2021rws,Roszkowski:2017nbc}, mirror matter~\cite{Foot:2014mia}, axions~\cite{Chadha-Day:2021szb,Aghaie:2024jkj}, and asymmetric dark matter (ADM)~\cite{Petraki:2013wwa,Hall:2021zsk}; axions are particularly compelling because they could simultaneously account for the DM population and solve the strong \textit{CP} problem in quantum chromodynamics~\cite{Kuster2008AxionsT}. 
Self-interacting dark matter (SIDM) offers a complementary avenue: by allowing DM particles to scatter off one another, SIDM models aim to reconcile small-scale structure tension, most notably, the core-cusp discrepancy in dwarf galaxy density profiles relative to collisionless $N$-body simulations~\cite{Spergel:1999mh}.

Recently, the study of NSs has increasingly been recognized as a leading indirect probe of DM. Massive stars can gravitationally accrete DM particles throughout their lifetimes \cite{Bertone:2007ae,Ilie:2020nzp,Robles:2022llu,Bose:2022ola,Koehn:2024gal}, and their core-collapse supernovae may synthesize additional DM in the collapsing core \cite{Janish:2019nkk,Chen:2022kal}. The resulting $\sim$\SI{10}{km} NS can therefore trap a substantial DM reservoir, these dark matter admixed neutron stars (DANS) thus providing a natural laboratory for testing DM physics under extreme conditions \cite{Luo:2025psd,Kumar:2025yei,Dengler:2025ntz}. 

DM captured by NSs introduces additional pressure-density components, modifying the total equations of state (EOS) and consequently altering stellar structure and observable properties (mass, radius, tidal deformability) \cite{Grippa:2024ach, Mukherjee:2025omu, Mahapatra:2024ywx}. These DM particles may also affect typical astrophysical processes inside NS, including the thermal evolution of NSs \cite{Ding:2019cky, Bhat:2019tnz, Avila:2023rzj} and multimessenger emissions from pulsars \cite{Ding:2019cky, Prabhu:2021zve, Suarez-Fontanella:2024epb}, enabling indirect constraints on DM particle properties through multimessenger observations \cite{Slatyer:2017sev}.
 
Initial theoretical frameworks for DANS centered on solving the two-fluid Tolman-Oppenheimer-Volkoff (TOV) equations for fermionic DM \cite{Leung:2011zz}, with subsequent studies exploring mass limit modifications \cite{Li:2012ii} and relativistic mean-field couplings \cite{Panotopoulos:2017idn}. Analytical approaches have derived parameter constraints directly from observational thresholds in both fermionic and bosonic models, including the $2M_\odot$ mass limit, gravitational wave (GW) and x-ray observations \cite{Sagun:2021oml, Karkevandi:2021ygv, Sagun:2022ezx, Mariani:2023wtv,Barbat:2024yvi}, as well as thermal evolution data from HESS~J1731-347 \cite{Sagun:2023rzp}. The field has recently progressed to Bayesian inference techniques that systematically combine those datasets from NSs to establish robust probabilistic constraints \cite{miao2022dark, Das:2020ecp, Rutherford:2022BosonicADM, Rutherford:2024uix, Arvikar:2025hej}. These methodological advances have yielded significant improvements in constraining the DM parameter space and collectively expanded our understanding of DANS.

Building upon these advances, the present work establishes a novel framework combining: (i) a two-fluid TOV formalism with $\alpha$-parametrized pressure anisotropy in the BM component, (ii) Bayesian inference of both BM and DM properties using multimessenger data, and (iii) systematic exploration of three representative BM EOS coupled with fermionic ADM. This approach uniquely addresses the limitations of previous studies by simultaneously incorporating anisotropic BM effects and consistent DM-BM interactions within a unified statistical framework. The physical motivation and implementation of this anisotropic pressure parameter $\alpha$ are detailed below.

Pressure anisotropy in NSs is characterized by the inequality between tangential and radial pressures: $P_t\neq P_r$. Several well-known microphysical processes in dense matter can produce anisotropies in NSs: the coupling to \textit{strong magnetic fields} of neutron spin \cite{Yazadjiev:2011ks}, anisotropic alignment of the pion field through \textit{Pion condensation} \cite{glendenning_1981}, \textit{superfluidity} of anisotropic $^3P_2$ pairing \cite{Hoffberg:1970vqj, Richardson:1972xn}, the \textit{tensor terms} in relativistic nuclear forces \cite{1975A&A....38...51H} and the core shear stresses in crystal \textit{phase transitions} \cite{Nelmes:2012uf}. Additional perspectives on these developments can be found in \cite{Mohanty:2023hha}.

Separately, a nonzero admixture of DM changes the stress-energy budget in a way that can imitate or mask this anisotropy \cite{Mahapatra:2024ywx}. Because baryonic anisotropy (parameter $\alpha$) and the DM mass fraction ($f_\chi$) imprint similar shifts on macroscopic observables such as the mass-radius curve and tidal deformability, treating them in isolation risks biased inferences. In the unified two-fluid Bayesian framework, both the anisotropy parameter $\alpha$ and the DM mass fraction $f_\chi$ are allowed to vary together. Fitting this model to multimessenger observations then reveals how much of a NS's structure is set by anisotropic BM versus captured DM.

The DM mass fraction 
$f_{\chi}$ is poorly constrained due to limited knowledge of stellar formation/accretion histories or DM properties (including mass and possible composition). However, permitting independent variations of $f_\chi$ for each star in our model would introduce excessive degrees of freedom, potentially leading to overfitting. To maintain tractability, we employ a shared $f_\chi$ across all DANSs. This simplification enables us to isolate the global impact of DM admixture on the mass-radius-tidal landscape \cite{Deliyergiyev:2023uer, DelPopolo:2020hel, Rutherford:2024uix}.

Since the high-density EOS of BM remains uncertain, we consider three well-established baryonic EOSs such as the Skyrme model, the relativistic Dirac-Brueckner-Hartree-Fock (DBHF) model, and a microscopic variational model, each extended to include anisotropic configurations in the presence of DM \cite{collier2022tidal}.  

Thanks to recent development from NASA's Neutron Star Interior Composition Explorer (NICER), NICER constrains NS's mass and radius from x-ray pulse-profile modeling, whereas GW170817 constrains tidal deformability from LIGO–Virgo gravitational-wave data. These are distinct measurements. We analyze them jointly by constructing a single multimessenger likelihood that incorporates both datasets. Earlier single-probe or fixed-EOS studies reported only best-fit EOS curves and thus provided no quantitative constraints on DM parameters. By jointly inferring the EOS band and propagating its uncertainty, our integrated approach yields statistically meaningful bounds on ADM mass, coupling, and fraction.  Key inputs include NICER measurements of PSR~J0030+0451 (J0030) \cite{Riley:2019yda,Miller:2019cac,vinciguerra2024updated}, PSR~J0740+6620 (J0740) \cite{miller2021,salmi2022radius,Dittmann24}, and PSR~J0437-4715 (J0437) \cite{choudhury2024nicer}, together with tidal deformability constraints from GW170817 reported by the LIGO Scientific Collaboration \cite{abbott2017gw170817,abbott2019tests,abbott2019properties,abbott2018gw170817}.

The paper is organized as follows: in Sec.~\ref{sec:theory}, we review the two-fluid TOV equations governing DANSs and describe the ADM model and the three BM EOSs employed. Sec.~\ref{sec:bayes} outlines the Bayesian inference methodology. Sec.~\ref{sec:TOV_acc} summarizes the difficulties encountered in solving the two-fluid TOV, and demonstrates the optimization to adapt it for Bayesian inference. Sec.~\ref{sec:analysis} presents and analyzes the macroscopic properties of DANS, the global constraints of NS parameters by posterior distribution, and phase transitions happening in the speed of sound inside NS. Finally, Sec.~\ref{sec:conclusion} offers concluding remarks and future directions.

\section{Theoretical Framework}\label{sec:theory}
\subsection{Equations of state}
This section lays out the EOS models adopted in our two-fluid framework, three benchmark BM EOSs for the baryonic sector and a Yukawa-interacting fermionic EOS for the DM sector, together with the assumptions required to merge them into a self-consistent model of admixed compact stars.
\subsubsection{Nuclear EOS}

The essential assumptions and properties of three different BM EOSs considered in this study are summarized in the following: a nonrelativistic Skyrme model, a relativistic DBHF model, and a microscopic variational model.

\begin{itemize}
    \item \textbf{Nonrelativistic Skyrme Model (Phenomenological Mean-Field Model)}: In this study, we employ the Brussels-Montreal functional BSk22 \cite{Pearson:2018tkr}, while details regarding BSk19-21 are available in Ref.~\cite{PhysRevC.88.024308}. These models utilize effective interactions, parameterized to reproduce experimental measurements, specifically the atomic masses of nuclei with $Z$, $N \geq 8$, based on the 2012 atomic mass evaluation \cite{Audi_2012}. The Hamiltonian framework encompasses terms dependent on density and other bulk BM properties. Notably, BSk22 is relatively stiff, enabling it to support massive NSs, consistent with observational constraints. The Hamiltonian describing the model is given by:
    \begin{align*}
    \mathcal{H}_{\text{Skyrme}} &= \frac{\hbar^2}{2m} \text{Tr} \left( \tau \nabla \psi \cdot \nabla \psi \right) 
    - \frac{1}{2} \text{Tr} \left( t_0 \rho^2 \right) \\
    &\quad - \frac{1}{6} \text{Tr} \left( t_3 \rho^3 \right)
    - \frac{1}{4} \text{Tr} \left( t_1 \left[ \nabla \rho \cdot \nabla \rho \right] \right) \\
    &\quad - \frac{1}{2} \text{Tr} \left( t_2 \left[ \rho \cdot \nabla^2 \rho \right] \right)~.
    \end{align*}
    Effective density-dependent terms with parameters $t_0, t_1, t_2, t_3$ are fit to nuclear data.
    \item \textbf{Relativistic DBHF Model}: The MPA1 EOS, characterized by its relative stiffness, is derived from the relativistic DBHF approach. Unlike nonrelativistic models such as Skyrme-based EOSs, MPA1 incorporates relativistic corrections and explicitly accounts for exchange interactions mediated by the $\pi$ and $\rho$ mesons \cite{Muther:1987xaa}.
     The Lagrangian density that captures the essential meson-nucleon interactions considered in the DBHF framework is given by:
    \begin{align*}
~~~~~~~~~~~~\mathcal{L_\text{DBHF}} &= \bar{\psi} \Big[ i\gamma^\mu \partial_\mu - M 
+ g_\sigma \sigma - g_\omega \gamma^\mu \omega_\mu \\
&\quad - g_\rho \gamma^\mu \bm{\tau} \cdot \bm{\rho}_\mu 
- \frac{f_\pi}{m_\pi} \gamma_5 \gamma^\mu \partial_\mu \bm{\pi} \cdot \bm{\tau} \Big] \psi \\
&\quad + \frac{1}{2} \left( \partial_\mu \sigma \, \partial^\mu \sigma 
- m_\sigma^2 \sigma^2 \right)  - \frac{1}{4} \omega_{\mu\nu} \omega^{\mu\nu} 
\\
&\quad+ \frac{1}{2} m_\omega^2 \omega_\mu \omega^\mu - \frac{1}{4} \bm{\rho}_{\mu\nu} \cdot \bm{\rho}^{\mu\nu} 
+ \frac{1}{2} m_\rho^2 \bm{\rho}_\mu \cdot \bm{\rho}^\mu  \\
&\quad+ \frac{1}{2} \left( \partial_\mu \bm{\pi} \cdot \partial^\mu \bm{\pi} 
- m_\pi^2 \bm{\pi} \cdot \bm{\pi} \right).
\end{align*}

    where the field tensors are defined as:
    \begin{align*}
\omega_{\mu\nu} &= \partial_\mu \omega_\nu - \partial_\nu \omega_\mu, \\
\bm{\rho}_{\mu\nu} &= \partial_\mu \bm{\rho}_\nu - \partial_\nu \bm{\rho}_\mu.
\end{align*}
   The DBHF approach is a relativistic extension of the Brueckner theory, where Dirac spinors describe nucleons, and the in-medium nucleon-nucleon interaction is obtained from a relativistic scattering matrix (G matrix). 
    
    \item \textbf{Microscopic Variational Model}: The AP3 EOS, based on a microscopic variational formalism, fundamentally differs from phenomenological mean-field models and relativistic DBHF approaches. It relies on realistic nuclear forces and rigorous many-body calculations, deriving the EOS from first-principles nucleon-nucleon interactions. Specifically, it utilizes the Argonne 18 potential while incorporating the $\delta$-meson ($\delta$) interaction but excludes the UIX three-nucleon potential, which results in a slightly stiffer EOS compared to AP4 \cite{Akmal:1998cf}. The Hamiltonian interpretation of this model is given by
    \begin{align*}
        \mathcal{H} &= \sum_{i} \frac{\mathbf{p}_i^2}{2m} + \frac{1}{2} \sum_{i \neq j} V_{ij} + \sum_{i < j < k} V_{ijk}~.
    \end{align*}
    Realistic interactions derived from two-body $ V_{ij} $ (e.g., Argonne V18) and three-body $ V_{ijk} $ (e.g., Urbana IX) potentials.
    
\end{itemize}

The three BM EOSs we consider capture the full landscape of BM behavior, from soft to stiff configurations. While MPA1 stands out as the stiffest model, supporting NSs up to \SI{2.48}{M_\odot}, AP3 and BSk22 follow with maximum masses of \SI{2.37}{M_\odot} and \SI{2.26}{M_\odot} respectively. This stiffness hierarchy interestingly reverses when examining typical \SI{1.4}{M_\odot} stars: BSk22 produces the most inflated configurations, MPA1 gives moderately compact stars, and AP3 yields the densest, most squeezed spheres of BM. Together, these models provide a physically motivated sampling of possible NS interiors. Allowing every EOS parameter to vary freely would make the model space dimension increase drastically and obscure the DM signal, whereas this compact set of EOS model choices keeps the analysis tractable while still capturing the key systematic uncertainty from the EOS side \cite{doi:10.1142/S021830131330018X}. To comprehensively assess uncertainties in the BM EOS within the relativistic-mean-field framework, we draw on Refs. \citep{Imam:2021dbe,Malik:2022ilb,Coughlin:2019kqf,Wesolowski:2015fqa,Furnstahl:2015rha,Ashton2019,Landry:2020vaw,CompactObject,Char:2023fue,Imam:2024gfh,2024MNRAS.529.4650H,2025MNRAS.536.3262H}. In the present work, we concentrate on exploring the impact of DM on NS properties.

\subsubsection{Fermions with Yukawa interactions}
To explore the effects of DM particles, we consider strongly self-interacting fermionic ADM with an EOS described by \cite{Kouvaris:2015rea, Mukhopadhyay:2016dsg}:

\begin{equation}
\label{eq:ferm}
    \begin{split}
       \rho_{\chi} =& \rho_{kin}(x) + \rho_{Y}(x) =  \frac{m_\chi^4}{8 \pi^2} \Bigl( x \sqrt{1+x^2} (1 + 2 x^2) \\
        &- \ln\left(x + \sqrt{1+x^2} \right) \Bigr) + \frac{g^2 x^6 m_\chi^6}{2 (3 \pi^2)^2 m_\phi^2} \\
        P_{\chi} =& P_{kin}(x) + P_Y(x)  =  \frac{m_\chi^4}{8 \pi^2} \Bigl( x \sqrt{1+x^2} \bigl( \frac{2 }{3} x^2 -1\bigr) \\
        &+ \ln\left(x + \sqrt{1+x^2} \right) \Bigr)  + \frac{g^2 x^6 m_\chi^6}{2 (3 \pi^2)^2 m_\phi^2}~,
    \end{split}
\end{equation}

where $\rho_{kin}(x)$ and $\rho_Y(x) $ are the kinetic energy density and Yukawa potential (arising due to DM self-interaction) energy density and $P_{kin}(x)$ and $P_Y(x) $ are the kinetic and Yukawa contributions to the pressure density, respectively. Here $ x = \frac{P_{F}}{m_\chi }$ is the dimensionless quantity (where $P_{F}$ denotes the Fermi momentum, $m_\chi$ is the mass of the DM fermion and we chose to work with natural units $\hbar = c =1$). The potential arising due to the self-interaction(repulsive) of two DM fermions $1$ and $2$ is considered to be of Yukawa type (repulsive in nature),
$V_{12} = \frac{g^2}{4 \pi} \frac{e^{-m_\phi r_{12}}}{r_{12}}$
where $r_{12}$ is the distance of separation of two DM fermions, 
$g$ is the Yukawa coupling constant (of the DM and the scalar mediator interaction term $g \bar{\chi} \chi \phi$), and $m_\phi$ is the mass of the scalar mediator i.e  $\chi \bar{\chi} \rightarrow \chi \bar{\chi}$ with $g$ at each vortex and  $\phi$ as a scalar mediator  as shown in the Feynman diagram with the self-interaction cross section $\sigma$ is given by

\begin{center}
    \begin{tikzpicture}
        \begin{feynman}
            \vertex (v1) at (-1,0) {};  
            \vertex (v2) at (1,0) {};   
            
            \vertex (a1) at (-2,1) {$  \bar{\chi}(p_2)$};
            \vertex (b1) at (-2,-1) {$\chi (p_1)$};
            \vertex (a2) at (2,1) {$\bar{\chi}(p'_2)$};
            \vertex (b2) at (2,-1) {$\chi(p'_1)$};
            \vertex (phi) at (0,0.5) {$\phi(q)$}; 
            
            \diagram* {
                (b1) -- [fermion] (v1) -- [fermion] (a1),  
                (b2) -- [fermion] (v2) -- [fermion] (a2),  
                (v1) -- [scalar, dashed] (v2), 
            };
            
            \filldraw (v1) circle (1.5pt);
            \filldraw (v2) circle (1.5pt);
            
            \node[left=6pt] at (v1) {$g$};
            \node[right=6pt] at (v2) {$g$};
        \end{feynman}
    \end{tikzpicture}
\end{center}
 
It is important to note that this EOS is consistent for repulsive self-interactions, which we consider. However, it may be inconsistent for relativistic fermions in the case of attractive self-interactions mediated by scalars (corresponding to the minus sign in the Yukawa contribution), which cannot become arbitrarily relativistic, since the pressure gradient must be positive to make the speed of sound satisfy the causality condition~\cite{WALECKA1974491, schmitt2014introductionsuperfluidityfieldtheoretical}.

The observed correlations between the EOS stiffness and the DM physical parameters $m_\chi$ (DM fermion mass) and $g$ (Yukawa coupling) stem from how these parameters influence the pressure contributions in the DM sector of the NS. As $m_\chi$ increases, the Fermi momentum $P_F \sim x m_\chi$ required to achieve the same number density $n_\chi$, also increases. But the dimensionless ratio $x$ decreases for this fixed $n_\chi$. Since both $P_{\text{kin}}(x)$ and $P_Y(x)$ are functions of powers of $x$, a smaller $x$ reduces both the pressure density components than the increase in energy density of DM EOS. As a result, the EOS becomes softer with increasing $m_\chi$, i.e., the pressure support at a given density is reduced. Conversely, increasing $g$ enhances the repulsive self-interaction between DM fermions, effectively adding a pressure term even at fixed number density $n_\chi$. Since $P_Y(x) \propto g^2$, a stronger $g$ boosts the overall pressure density, particularly in the intermediate density regime, where the interaction terms are non-negligible but not yet dominated by relativistic effects. This results in a stiffening of the EOS.

\begin{figure}[htbp]
    \centering
    \includegraphics[width=\columnwidth]{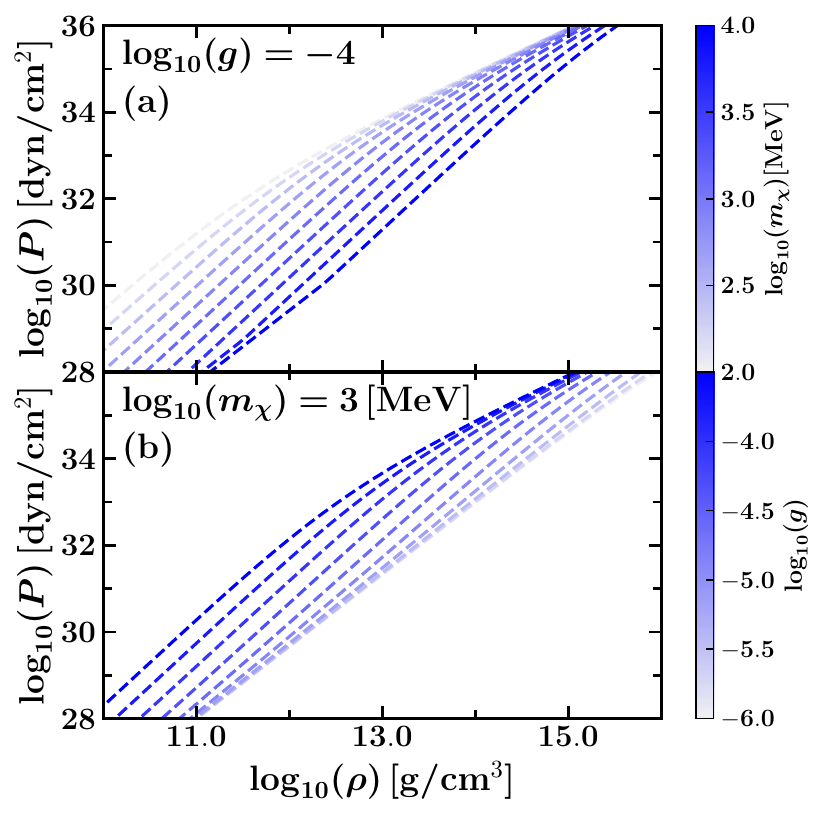}
    \caption{\justifying The DM EOS as a function of $m_\chi$ (upper) and $g$ (lower) while fix $\log_{10}g=-4$ ($m_\chi=\SI[parse-numbers=false]{10^3}{MeV}$)}
    \label{fig:EOS_ex}
\end{figure}

Figure~\ref{fig:EOS_ex} (a) demonstrates the softening effect on the DM EOS when increasing $m_\chi$ while holding $g$ fixed at $10^{-4}$. Here, the ten dashed pressure-density ($P$-$\rho$) curves exhibit systematic downward shifts with growing $m_\chi$ and deeper colors. Conversely, Figure~\ref{fig:EOS_ex} (b) shows the EOS stiffening behavior when $m_\chi$ is fixed at $\SI[parse-numbers=false]{10^{3}}{MeV}$ and $g$ is increased. In this case, the $P$-$\rho$ relations display upward trends, indicating enhanced pressure support at given densities. This explains the observed inverse correlation between EOS stiffness and $m_\chi$, and the positive correlation with $g$, within the constraints imposed by observationally allowed self-interaction cross sections ($\sigma/m_\chi \sim 0.1-10 \,\text{cm}^2/\text{g}$) \cite{Markevitch:2003at, Kaplinghat:2015aga}.

\subsection{TOV framework}

The two-fluid computational framework employed in this work is based on the open-source code developed by \cite{collier2022tidal,michael_collier_2022_7361819}\footnote{\url{https://doi.org/10.5281/zenodo.7361819}}, which has been further enhanced to include the treatment of anisotropic effects and to improve the computational efficiency for this study to enable Bayesian inference.

Assuming negligible interaction between BM and DM, we treat them as separate fluids governed by the following coupled structure equations. The total energy density and radial pressure are given by $\rho = \rho_B + \rho_\chi$ and $P_r=P_{rB}+P_{r\chi}$, respectively. Each fluid satisfies its own generalized TOV equation,
\begin{equation} \label{eq:TOVi}
    \frac{dP_{ri}}{dr} = -\frac{(\rho_i + P_{ri})(m + 4\pi r^3 P_r)}{r(r - 2m)} + \frac{2\Pi_i}{r},
\end{equation}
where $i=B,\chi$ refers to BM or DM, and $ \Pi_i = P_{ti} - P_{ri} $ characterizes the pressure anisotropy of each component. We assume the DM component to be isotropic, i.e., $P_{t\chi}=P_{r\chi}$, and thus $ \Pi_\chi = 0 $. Anisotropy is introduced only in the baryonic sector via the Bowers–Liang (BL) model \cite{1974ApJ...188..657B, Das:2022ell}, where the tangential pressure $P_{tB}$ is given as
\begin{align} \label{eq:anisoBL}
    P_{tB} = P_{rB} + \frac{\alpha}{3} \dfrac{(\rho_B + 3 P_{rB}) (\rho_B +  P_{rB}) r^2}{1 - 2m / r}.
\end{align}
Here, the parameter $\alpha$ controls the degree of anisotropy, and setting $\alpha = 0$ restores isotropy in the baryonic sector as well. In general, positive values of $\alpha$ systematically generate larger masses and radii for a given central density. This well-established relationship, discussed in \cite{Das:2022ell} and references therein, persists in DANS as demonstrated in our previous work \cite{Mahapatra:2024ywx}.

The mass $m_i$ enclosed within radius $r$ satisfies:
\begin{equation} \label{eq: dmdri}
\frac{dm_i}{dr} = 4\pi r^2 \rho_i.
\end{equation}

Boundary conditions ensure pressure vanishes at $R_B$ and DM core radius $R_\chi$,
\begin{equation} \label{eq:pR}
    P_i(R_i) = 0.
\end{equation}

Two possible BM-DM configurations exist
\begin{itemize}
\item\textbf{DM confined to the NS core}: $R_B > R_\chi$.
\item\textbf{DM forming a halo}: $R_\chi > R_B$.
\end{itemize}

The observed BM radius is identified as $R_B$, while total mass $M$ and DM fraction $f_\chi$ are given by:
\begin{equation} \label{eq:Mt}
    M = M_B(R_B) + M_\chi(R_\chi),
\end{equation}
\begin{equation} \label{eq:fchi}
    f_\chi = \frac{M_\chi(R_\chi)}{M}.
\end{equation}
By varying central densities, we obtain different combinations of $R_B$, $R_\chi$ and $M$ for a given $f_\chi$.

When solving the two-fluid TOV equations, the initial conditions require precise specification of the central DM density $\rho_{\chi,c}$. One can get this either with the global mass fraction $f_\chi = M_\chi/M$ or with the central-density ratio $f_{c,\chi} = \rho_{c,\chi}/\rho_c$.  Although the latter integrates slightly faster, we adopt $f_\chi$ for the Bayesian runs because it has a transparent astrophysical meaning, set by the star’s cumulative DM capture, and yields tighter, more data-driven posteriors.  In contrast, $f_{c,\chi}$ is tied to the uncertain microscopic DM EOS, spreads the sampler over a much broader volume, and dilutes the constraining power of the observations.

The radius inferred from observations is not simply the BM radius $R_B$.  In a dark-halo scenario, photons leaving the stellar surface experience extra red-shift and light-bending, so NICER’s pulse-profile fits would overestimate the true stellar size.  Current NICER errors still dominate that bias \cite{Shawqi:2024jmk}, but next-generation x-ray missions will have to model it explicitly.  An analogous issue appears in binary-neutron-star mergers: the two DM halos merge first, modifying the early tidal field and thus the tidal deformability we inferred from GW data.

To remain agnostic about halo size, we follow \cite{collier2022tidal,michael_collier_2022_7361819} and define the observable radius $R$ as the sphere that encloses 99.99 \% of the total mass:
\begin{equation}
    \int_0^R 4\pi r^2 \left[\rho_B(r) + \rho_\chi(r)\right] dr = 99.99\%\times M~.
\end{equation}
Scanning over the central densities of both fluids then maps the full set of allowed mass-radius pairs $(M, R)$ for any chosen DM fraction $f_\chi$.

\subsection{Tidal deformability} \label{tidef}

We follow the standard relativistic-perturbation formalism
\cite{Hinderer_2010,Damour_2009,Flanagan_2008} and extend it to an
anisotropic two–fluid system as in
\cite{Mahapatra:2024ywx}.  The dimensionless tidal deformability
remains
\begin{equation}
      \Lambda = \frac{2}{3}\,k_2\,C^{-5}, 
  \qquad
  C \equiv \frac{M}{R},
\end{equation}

with the Love number $k_2$ given by the usual algebraic function of the
surface variable $y_R$ (see \cite{Damour_2009}).

The perturbation equation keeps its form
\begin{equation}
      r y' + y^2 + (r M(r) - 1)\,y + r^2 N(r) = 0 ,
\end{equation}
but the source terms $M(r)$ and $N(r)$ now includes the \emph{sum} of BM and DM
contributions to density, pressure, and sound speeds, plus the
BL anisotropy. Explicit expressions are in 
\cite{Mahapatra:2024ywx}.  Boundary conditions are unchanged:
$P(0)=P_c$, $m(0)=0$, $y(0)=2$.

In practice we integrate this $y$-equation alongside the two-fluid TOV system, and then evaluate $k_2$ and $\Lambda$ at $R$. Numerical details follow \cite{collier2022tidal,michael_collier_2022_7361819}.
\section{Bayesian Methodology}\label{sec:bayes}
\subsection{Choice of priors for model parameters} \label{sec:prior}

The DM EOS is characterized by three microphysical parameters: the fermion mass $m_\chi$, the scalar mediator mass $m_\phi$, and the Yukawa coupling $g$ [Eq.~\eqref{eq:ferm}]. Current accelerator \cite{Alexander:2016aln}, direct detection \cite{XENON:2018voc, DarkSide:2018bpj}, and Bullet-Cluster self-interaction bounds \cite{Randall:2007ph, Tulin:2017ara} still permit fermionic ADM to have subweak Yukawa couplings while accumulating in NSs; see also the reviews \cite{Petraki:2013wwa, Kouvaris:2015rea} for astrophysical implications. Guided by those limits, we adopt log-uniform priors:
$
\SI[parse-numbers=false]{10^{2}}{MeV} < m_\chi < \SI[parse-numbers=false]{10^{4}}{MeV},
\SI[parse-numbers=false]{10^{-6}}{} < g < \SI[parse-numbers=false]{10^{-3.5}}{},
$
spanning the canonical $\mathcal{O}(\mathrm{GeV})$ ADM mass scale and reaching the Bullet-Cluster ceiling on velocity-independent self-interaction cross sections.  We fix the mediator mass to $m_\phi = 10^{-3}\,\text{MeV}$ so that the interaction range matches typical interparticle separations in NS cores while avoiding resonant self-capture. 

For all three BM EOS considered in this work, we adopt identical $\alpha$ parameter ranges to characterize BM anisotropy. Specifically, we impose a uniform prior distribution from -2 to 2, sufficiently broad to encompass the stable, causality-respecting values predicted for magnetic, superfluid, or phase-transition driven anisotropies without venturing into mechanically unstable regimes.

In the two–fluid description, we parametrize the DM content by the mass fraction $f_\chi \equiv M_\chi/M_T$. Standard capture calculations give $f_\chi \ll 1\%$, yet exotic production or collapse channels could boost it to the few percent level.  To stay agnostic we impose a broad prior $0 < f_\chi \le 10\%$. The nested-sampling engine we implemented here cannot reliably explore values below $f_\chi \simeq 10^{-6}$; this numerical floor is still 5 orders of magnitude below the percent-level admixtures that dominate the observable effects in our analysis. 

The complete prior distributions for all model parameters are summarized in Table~\ref{tab:prior}.
\begin{center}
    \begin{table}
    \renewcommand{\arraystretch}{1.5}
    \setlength{\tabcolsep}{15pt}
        \begin{tabular}{cc}
            \hline
            \text{\textbf{EOS parameter}} & \textbf{Prior} \\
            \hline
            $f_\chi$ & $\mathcal{U}(0,\, 10\%)$ \\
            $m_{\chi}$ [MeV] & $\mathcal{LU}(10^2,\, 10^4)$ \\
            $g$ & $\mathcal{LU}(10^{-6},\, 10^{-3.5})$ \\
            $\alpha$ & $\mathcal{U}(-2,\, 2)$ \\
            $\rho_c$ [MeV$^4$] & $\mathcal{LU}[10^{9.1}, \max(\rho_{c,\mathrm{\max}},\, 10^{10.8})]$ \\
            \hline
          \end{tabular}
          \caption{Summary of prior settings for inference parameters. $\mathcal{U}$ denotes a uniform (flat) distribution. $\mathcal{LU}$ denotes a log-uniform distribution.}
          \label{tab:prior}
        \end{table}
    \end{center}

\subsection{Bayesian framework}
The Bayesian inference methodology employed in this study adheres to the framework established and utilized in the works of \cite{greif2019equation}, as well as \cite{raaijmakers2019nicer,raaijmakers2020constraining,raaijmakers2021constraints,Huang:2023grj,Huang:2024rfg,Huang:2024rvj,Huang:2025vfl}. According to Bayes' theorem, the posterior distribution of $\boldsymbol{\theta}$ follows from multiplying its prior by the likelihood and normalizing with the Bayesian evidence $p(\boldsymbol{d}\,|\,\mathbb{M})$, where $\boldsymbol{d}$ denotes the observational dataset, and $\mathbb{M}$ represents all the physical and mathematical models we consider here:
\begin{equation} \label{eq:Bayes}
{
p(\boldsymbol{\theta} \,|\, \boldsymbol{d},\, \mathbb{M}) =
\frac{
    p(\boldsymbol{\theta}\,|\,\mathbb{M})
    p (\boldsymbol{d}\,|\,\boldsymbol{\theta},\, \mathbb{M})
}{
    p(\boldsymbol{d}\,|\,\mathbb{M})
}
}.
\end{equation}
Prior $p(\boldsymbol{\theta}\,|\,\mathbb{M})$ is empirically determined as summarized in Sec.~\ref{sec:prior}, the likelihood $p (\boldsymbol{d}\,|\,\boldsymbol{\theta}, \mathbb{M})$ quantifies observation probabilities given model predictions. However, in our case, the constraints of likelihood are derived from independent observations of NSs, therefore, we need to take the central density $\boldsymbol{\rho_c}$  into account to match each observational result and transform Eq. (\ref{eq:Bayes}) to   
\begin{equation}
p(\boldsymbol{\theta},\, \boldsymbol{\rho_c}\,|\,\boldsymbol{d},\, \mathbb{M}) \propto  
p(\boldsymbol{\theta}\,|\,\mathbb{M}) p (\boldsymbol{\rho_c}\,|\,\boldsymbol{\theta},\,\mathbb{M}) 
p(\boldsymbol{d}\,|\,\boldsymbol{\theta},\, \boldsymbol{\rho_c},\, \mathbb{M}).
\end{equation}
The likelihood function is divided into two parts because the permissible range of $\boldsymbol{\rho_c}$ for stable NSs is determined by $\boldsymbol{\theta}$. Here we set the lower bound of $\boldsymbol{\rho_c}$ to $\SI[parse-numbers=false]{10^{9.1}}{MeV^4}$ (around $\SI[parse-numbers=false]{10^{14.5}}{g/cm^3}$), while the upper bound takes the larger value between $\boldsymbol{\rho_{c,\mathrm{max}}}(\boldsymbol{\theta})$ and $\SI[parse-numbers=false]{10^{10.8}}{MeV^4}$ (around $\SI[parse-numbers=false]{10^{16.2}}{g/cm^3}$), where $\boldsymbol{\rho_{c,\mathrm{max}}}$ represents the $\boldsymbol{\rho_{c}}$ at the NS's maximum mass configuration.

In this work, weighted sampling of the parameter vector $\boldsymbol{\theta}$ is accomplished by the nested sampling Monte Carlo algorithm MLFriends \cite{buchner2016statistical,buchner2019collaborative} using the \texttt{UltraNest}\footnote{\url{https://johannesbuchner.github.io/UltraNest/}} package \cite{buchner2021ultranest}. The same sampling algorithm has been implemented and documented in the open-source \texttt{CompactObject} package, a Bayesian inference framework to constrain the NS EOS by jointly analyzing astrophysical observations and nuclear-experimental data \citep{CompactObject,cartaxo2025completesurveytextttcompactobjectperspective}.

In this analysis, we utilized the $M$ and $R$ posteriors obtained from pulse profile modeling of observations of three-millisecond pulsars: J0030 \cite{vinciguerra2024updated}, J0740 \cite{salmi2022radius}, and J0437 \cite{choudhury2024nicer}. Additionally, we incorporated the tidal deformability measurement derived from the binary NS merger event GW170817 \cite{abbott2017gw170817, abbott2019tests, abbott2019properties}, as detected by LIGO.  The overall likelihood function could be written as
\begin{align*}
  p (\boldsymbol{d}\,|\,\boldsymbol{\theta},\, \boldsymbol{\rho_c},\, \mathbb{M})
  & = p (\boldsymbol{d} \,|\,
  \mathbf{M}, \mathbf{R},\, \Lambda,\, \mathbb{M})\\
  & \propto \prod_i p (\boldsymbol{d}_{\mathrm{NICER}, i}\,|\,M_i,\, R_i,\, \mathbb{M}) \\
  & \times p (\boldsymbol{d}_{\mathrm{GW}}\,|\,\Lambda_1,\, \Lambda_2,\, M_1,\, M_2,\,\mathbb{M})\\
  & \propto \prod_i p (M_i,\, R_i\,|\,\boldsymbol{d}_{\mathrm{NICER},i},\, \mathbb{M}) \\
  & \times p (\Lambda_1,\, \Lambda_2,\, M_1,\, M_2\,|\,\boldsymbol{d}_{\mathrm{GW}},\, \mathbb{M}).
\end{align*}

Moreover, since the chirp mass is measured with exceptional precision, we have adopted the methodology proposed by \cite{raaijmakers2020constraining,Huang:2023grj} fixed it to $M_{\text {c }}=\left(M_{1} M_{2}\right)^{3 / 5}/\left(M_{1}+M_{2}\right)^{1 / 5}$ to the median value $M_{\text {c}}=1.186$ M$_{\odot}$ ~for GW170817. Specifically, we transform the posterior distributions of the binary event into those of the two tidal deformability parameters $\Lambda_1$ and $\Lambda_2$, the mass ratio $q$, and the chirp mass $M_\mathrm{chirp}$. We can write down the likelihood part of GW170817 as
\begin{align}
    p (\Lambda_1,\, \Lambda_2,& M_1, M_2\,|\,\boldsymbol{d}_{\mathrm{GW}},\, \mathbb{M}) \nonumber\\
    & \sim p (\Lambda_1,\, \Lambda_2, q\,|\,\boldsymbol{d}_{\mathrm{GW}},\, \mathbb{M}_\mathrm{chirp}).
\end{align}
To accelerate convergence in our Bayesian inference, we adopt the mass ordering constraint $M_1 > M_2$ for gravitational wave events, consistent with \cite{raaijmakers2020constraining,Huang:2023grj}.

To guide a better convergence direction for the sampler, we further implement a \emph{graded-penalty} filtering scheme. The criterion uses a physically motivated cutoff $R_\text{cut} = \SI{40}{km}$: any $(M, R)$ curve with minimum radius $R_\mathrm{min}$ exceeding this value would produce NSs too large to be consistent with multi-messenger observations (formally corresponding to Liklihood $\mathbb{L}$: $ \log(\mathbb{L})\ll -10^{5}$). For such cases, we skip the computationally expensive two-fluid TOV integration by assigning empirical surrogate values $(M_{\mathrm{sur}}, R_{\mathrm{sur}})$ to the $(M, R)$ pairs, where these surrogate quantities are defined through the following prescription: 
\begin{equation} \label{surrogate_radius}
\begin{aligned}
    M_{\mathrm{sur}} &= M(R_{\min})\times
        \frac{\log_{10}(\rho_c)-\log_{10}(\rho_{\min})}{\log_{10}(\rho_{\max})-\log_{10}(\rho_{\min})},\\
    R_{\mathrm{sur}} &= R_\mathrm{min}
    \left[1+2\,\frac{\log_{10}(\rho_{\max})-\log_{10}(\rho_c)}{\log_{10}(\rho_{\max})-\log_{10}(\rho_{\min})}\right],\\
\end{aligned}
\end{equation}
 a linear map between the bounds $\log_{10}(\rho_{\min} )$, $\log_{10}(\rho_{\max})$  on $\log_{10}(\rho_c)$. This mapping approximates the true $(M, R)$ curve and smoothly down-weights the likelihood as $R_{\min}$ grows, so the nested sampler still feels a gradient that points back toward physically admissible configurations while avoiding wasted compute in clearly unphysical regions. In our prior space, we verified that the filter triggers almost exclusively in the initial iterations (before live points reach the high-likelihood region). It cuts compute time but does not change the results like posterior distribution, the inferred EOS band, and the log-evidence match runs without the filter within Monte Carlo uncertainty.

\section{Optimizing the Two-Fluid TOV Solver}\label{sec:TOV_acc}
With the parameter ranges fixed, \texttt{UltraNest} samples the EOS space and, for each sample, calls the TOV solver, computes the model observables, evaluates their likelihood against the data, and updates the posterior until convergence. We configured \texttt{UltraNest} with $4000$ live points and terminated sampling upon achieving 99\% parameter space exploration. This resulted in $\sim 46,000$ CPU hours and $\sim 70,000$ iterations per DANS model. 

Prior to implementing the Bayesian analysis, however, we need to perform some critical optimizations on the TOV code framework originally presented in \cite{collier2022tidal,michael_collier_2022_7361819} to adapt it for Bayesian inference. Specifically, in the two-fluid TOV framework calculating $M(\rho_c)$ requires solving an inherent inverse problem: while EOS parameters, $f_\chi$, and $\rho_c$ are specified, the initial condition $\rho_{c,\chi}$ remains unknown. Leveraging the monotonic relationship between $\rho_{c,\chi}$ and $f_\chi$, \cite{collier2022tidal,michael_collier_2022_7361819} adopted a bisection search over $\rho_{c,\chi} \in [0, \rho_c]$ to iteratively match the target $f_\chi$ by solving the TOV system repeatedly. This nested procedure incurs substantial computational overhead compared to single-fluid TOV or simplified two-fluid models with fixed $f_{c,\chi}$. In Bayesian runs we evaluate many EOS models. We use those results to give the bisection that finds $\rho_{c,\chi}$ a much better starting bracket. The scheme has three layers:

\begin{itemize}
\item \textbf{Precomputed map (one-time):}
We build a lookup table on a $13^5=371{,}293$-point grid over $(m_\chi, g, f_\chi, \alpha, \rho_c)$ and store the corresponding $\rho_{c,\chi}$ solutions. This serves as a global reference that we can reuse throughout sampling.

\item \textbf{Setting tight bounds :}
(i) \textit{Nearest-neighbors bracket:} For any new parameter set, we find the $3^5=243$ closest grid points and take the minimum and maximum of their stored $\rho_{c,\chi}$ as the initial lower/upper bounds. This typically cuts the number of bisection steps by about a factor of 3.\\
(ii) \textit{On-the-fly interpolation:} After we have solved at least five $(\rho_c,\rho_{c,\chi})$ pairs for the \emph{same} EOS parameters $(m_\chi, g, f_\chi, \alpha)$, we fit a straight line to $\rho_{c,\chi}$ versus $\rho_c$ and use the predicted value (with a small safety margin) to set even tighter bounds for the next call. This speeds up the solve for off-grid points.

\item \textbf{Safe fallback if a guess is poor:}
If a guessed bracket does not contain the solution, we first widen it by 20\% and try again. If that still fails, we reset to a conservative default range $[0,\rho_c]$. This prevents crashes and guarantees progress.
\end{itemize}

Furthermore, beyond these optimizations on the TOV solving process, we enhanced the determination of $\rho_{c,\max}$ by refining our team’s original approach from the \texttt{CompactObject} v2.0 package. The conventional single-fluid implementation relied on uniform sampling of $50$ points across the $\log_{10}(\rho_c)$ prior space to identify the maximum mass point. For the two-fluid case, where computational costs are substantially higher, we implemented a unimodal peak-finding routine based on Kiefer’s golden-section search algorithm \cite{kiefer1953sequential}.

\section{Analysis Results}\label{sec:analysis}

Running separate Bayesian analyses with the three benchmark EOS: BSk22, MPA1, and AP3, augmented by anisotropic stresses, admixed with the DM model, lets our DANS model sweep the full soft-to-stiff landscape of BM. In the following four subsections, we will provide a detailed discussion of our key findings.

\subsection{Global constraints on DANS}\label{sec:GlobalDANS}

\begin{figure*}[htbp]
    \centering
    \includegraphics[width=1\textwidth]{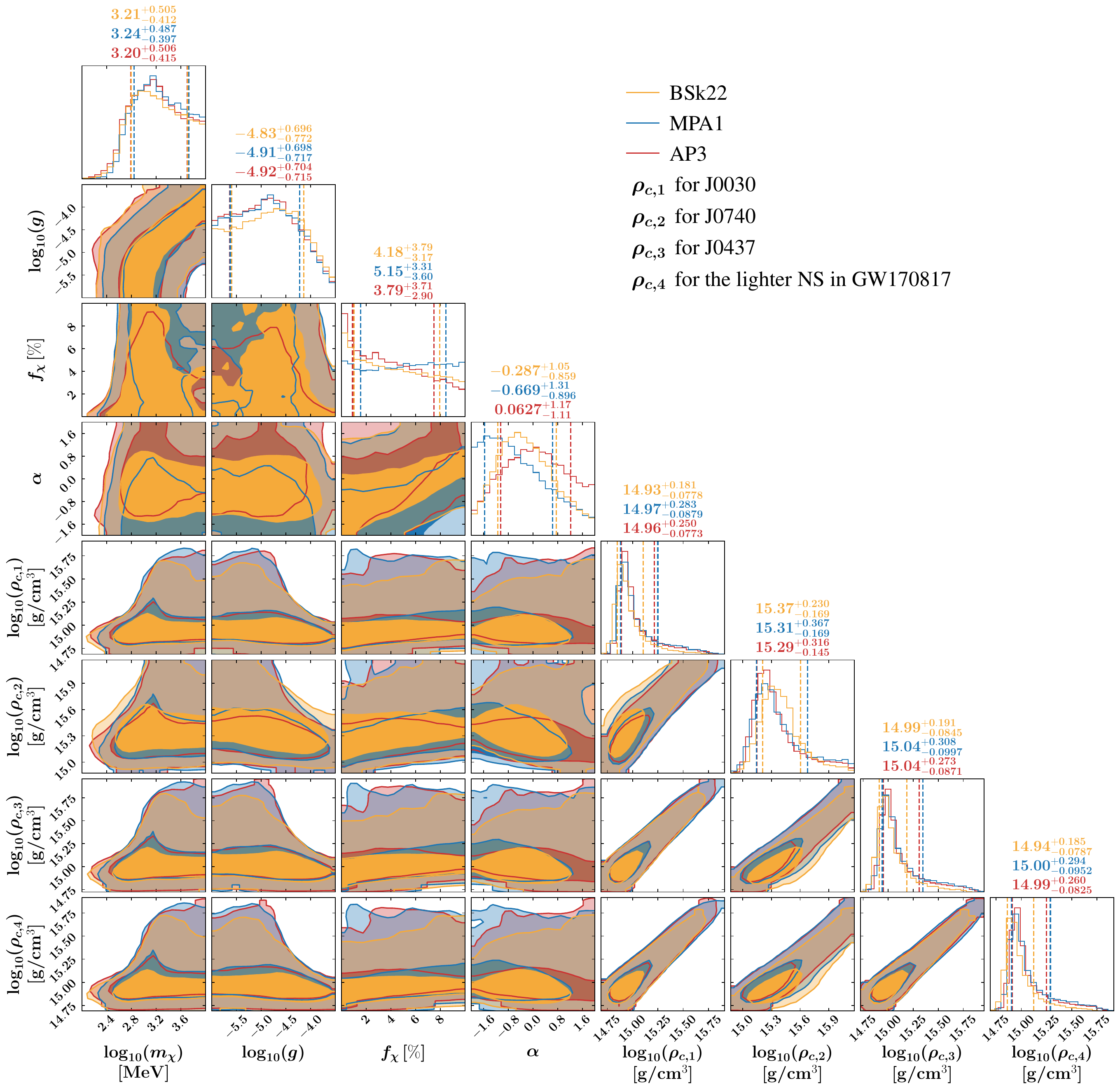}
    \caption{\justifying Corner plot showing the 1D marginalized posteriors and 2D joint distributions of DM properties, anisotropy parameter, and central densities in the DANS framework, obtained via Bayesian inference using combined x-ray and GW data. The densities labeled from $\rho_{c,1}$ to $\rho_{c,4}$ correspond to the central densities of J0030, J0740, J0437, and the lighter NS in GW170817, respectively. Color-coded contours represent DM admixed with different BM EOS: BSk22 (yellow), MPA1 (blue), and AP3 (red). Darker and lighter regions correspond to 68\% and 99\% credible regions respectively. In the 1D corner plots, the two dashed lines mark the 15.87th and 84.13th percentiles of the parameter distributions. The titles display the median value along with these percentile bounds for each parameter.}
    \label{fig:param_all}
\end{figure*}

\begin{figure}[htbp]
    \centering
    \includegraphics[width=\columnwidth]{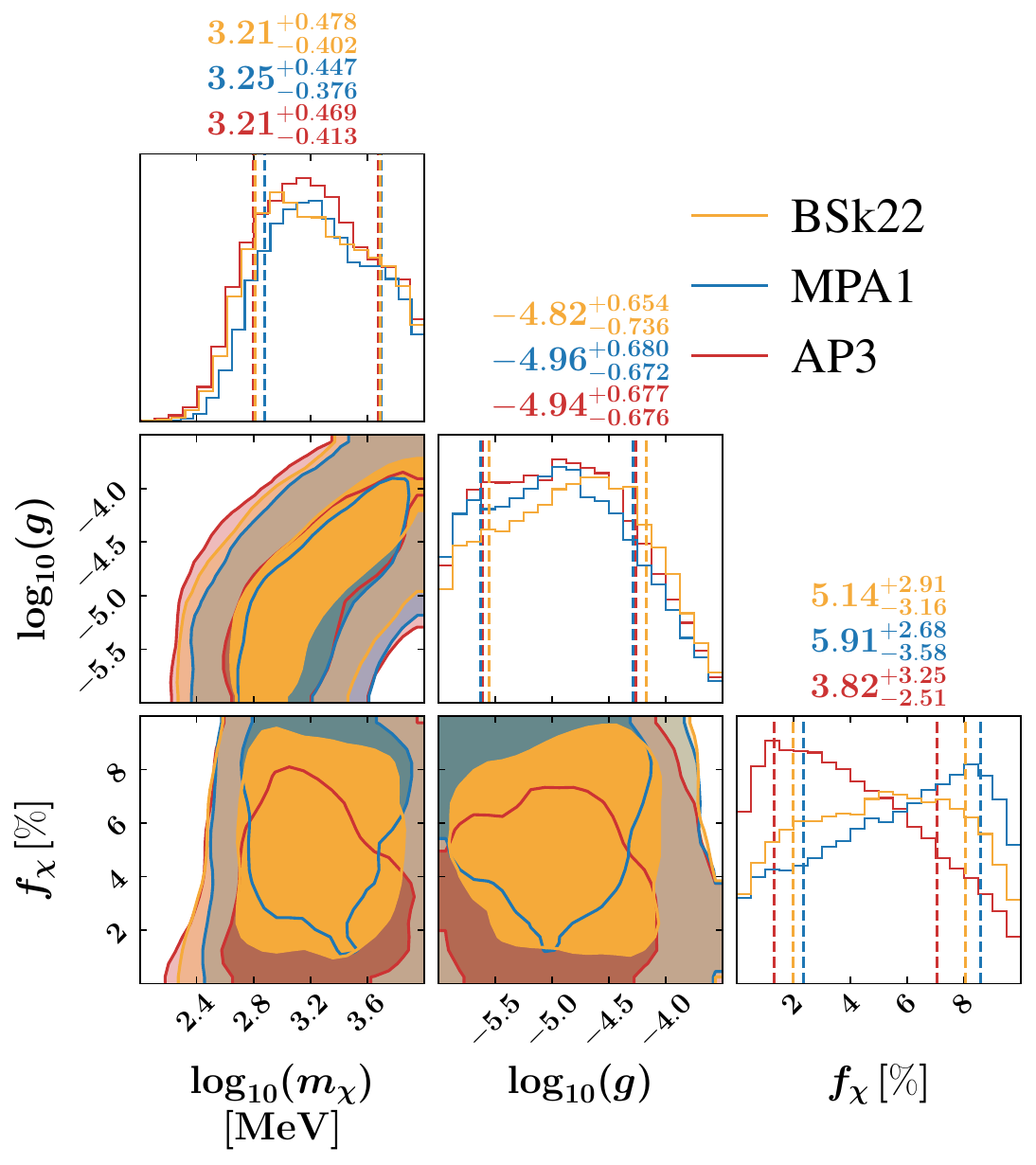}
    \caption{\justifying The posterior distributions of DM parameters under the isotropic BM constraint ($|\alpha|< 0.001$), derived from the resampling of the full DANS parameter space. The shaded contours, dashed lines, and numerical annotations follow the same conventions as described for Figure~\ref{fig:param_all}}
    \label{fig:fixed_alpha}
\end{figure}

Figure~\ref{fig:param_all} presents the marginalized posterior\footnote{See the Zenodo repository \cite{liu_2025_17082249} for the data and code to recreate all plots in this paper.} distributions for the key parameters governing the DANS. These parameters include the DM particle mass $m_\chi$, the coupling constant $g$ between DM and its mediator $m_\phi$, the DM fraction in DANS $f_\chi$, the anisotropy parameter $\alpha$, and the set of central energy densities $\{\rho_{c,1},...,\rho_{c,4}\}$ corresponding to different DANS configurations. $m_\chi$, $g$, $f_\chi$, and $\alpha$ are shared parameters of all observations, while the densities from $\rho_{c,1}$ to $\rho_{c,4}$ correspond to the central densities of J0030, J0740, J0437, and the lighter NS in GW170817, respectively. A clear departure from the uniform prior distributions for most of the DANS EOS parameters demonstrates that current observational data already effectively constrain these parameters, with the posteriors revealing distinct correlations and degeneracies among them.

The posteriors for $m_\chi$ and $g$ "appear" to be strongly constrained: they favor $(\text{GeV})$ scale fermion masses, remain well below weak coupling, and fall along a broken straight line in the $\log_{10}(m_\chi)$–$\log_{10}(g)$ plane, just as seen in the core-only analysis of \cite{Rutherford:2024ADM}. This line, however, is not carved out by the data but by our modeling choice to define the stellar radius as the sphere enclosing $99.99\%$ of the total mass.  This definition systematically rejects configurations with extended DM halos, causing the sampler to preferentially explore DM-core solutions that \emph{must} lie on the same broken track. Sec.~\ref{sec:mac_prop_rx} will discuss this model preference in more detail and demonstrate how little the present observation data can say something about the DM sector.

Comparing different EOS models resulted in a posterior constraint; the central value of the $\alpha$ posterior distribution heavily depends on the stiffness of BM EOS: while MPA1 supports the largest maximum mass ($\SI{2.4}{M_\odot}$) with the lowest $\alpha$ peak, AP3 demonstrates its softer nature by producing more compact configurations (smaller radii) below $\SI{2}{M_\odot}$, corresponding to higher $\alpha$ values. The inferred values of $\alpha$ in DANS are systematically bigger than those obtained for BM-only NSs from Bayes analysis (see Appendix, Figure \ref{fig:param_alpha_all}). In the inference that includes only BM, the data often drive $ \alpha $ posterior toward the edges of its prior range to reproduce the observed masses and radii.  Introducing a DM component alters the global pressure balance, so equally good fits can be obtained without large anisotropic corrections. The posterior for $ \alpha $ therefore contracts toward moderate values ($|\alpha|\lesssim 0.5$), indicating that strong anisotropy in BM is no longer required once DM is present.

The generally flat $f_\chi$ distributions indicate relatively weak constraints on the DM mass fraction in DANS, highlighting the need for improved observational sensitivity to DM contributions. Nevertheless, subtle EOS-dependent differences emerge: while BSk22 and AP3 show mild preferences for smaller $f_\chi$ values, MPA1 exhibits the flattest distribution. This behavior correlates with MPA1's tightly constrained $\alpha$ distribution, whose peak concentrates near the prior boundaries. The combined trends reveal that softening the stiff MPA1 requires both reducing the value of $\alpha$ and compensating DM contributions to overcome the limitations imposed by the lower $\alpha$ boundary—-a requirement less pronounced for the softer BSk22 and AP3.

As the anisotropic environment plays a similar role to DM, we performed a resampling of the posterior distribution in Figure~\ref{fig:fixed_alpha} to isolate the impact of DM from anisotropic effects. By using Gaussian kernel estimation, we select the posterior samples that satisfy the condition $|\alpha|< 0.001$. Figure~\ref{fig:fixed_alpha} highlights two main results that hold across all three models.  
First, when $\alpha$ is restricted to around zero, the posterior distributions of $f_\chi$ deviate noticeably from its flat prior regardless of the choices of BM EOS; the position of its peak depends on the stiffness of BM EOS. This behavior confirms that the $f_\chi$ posterior depends on the determination of BM EOS, although for $\alpha\neq0$ this dependence is largely washed out by anisotropy, producing an almost flat posterior.  

Second, the posterior distributions of $m_\chi$ and $g$ show little variation among different BM EOS choices, in the nearly isotropic limit. Because $m_\chi$ and $g$ are set mainly by the fundamental properties of the DM sector of the EOS, current multimessenger observations do constrain these two parameters, yet that very insensitivity to the BM EOS choices means their posteriors alone cannot distinguish between different BM models.

The four central density distributions in the bottom-right panel of Figure~\ref{fig:param_all} display remarkable consistency and positive correlations. Compared to the BM-only cases in the Appendix (Figs.~\ref{fig:param_alpha_all} and \ref{fig:param_EOS}), $\rho_{c,i}$ of DANS exhibit three distinct features: (\romannumeral1) broader distributions, (\romannumeral2) reduced sensitivity to BM composition, and (\romannumeral3) systematically higher medians. The third feature directly demonstrates DM's softening effect: the reduced pressure support requires larger $\rho_c$ to sustain equivalent stellar masses. Meanwhile, the first two features reflect how the gravitationally coupled DM component introduces greater uncertainty in density determination and obscures BM composition differences. As for the first feature, the physical origin of the extended distributions of $\rho_{c,i}$ (68\%--99\% credible intervals) will be thoroughly examined in Sec. \ref{sec:mac_prop_rx}.

While the inclusion of DM introduces three additional parameters and presents challenges in constraining DM parameters across BM models (Figs.~\ref{fig:param_all}, \ref{fig:fixed_alpha}), one might question whether such complexity is justified. To quantitatively address this query, we extract the Bayesian evidence directly from the posterior output files of \texttt{UltraNest} and compute the Bayes' factor \cite{kass1995bayes} between the evidences of two models. This analysis allows for a quantitative comparison of different models to determine if the data favor one model over another.

\begin{table}[htbp]
\begin{threeparttable}
\setlength{\tabcolsep}{10.5pt}
\begin{tabular}{lccc}
\toprule
 & \multicolumn{3}{c}{EOS} \\ 
 \cmidrule(lr){2-4}
 Configuration & \textbf{BSk22} & \textbf{MPA1} & \textbf{AP3} \\ 
\midrule
\textbf{DANS}         & $-17.8(1)$ & $-17.8(1)$ & $-18.2(1)$ \\
\textbf{Anisotropic}  & $-16.8(1)$ & $-17.2(1)$ & $-17.2(1)$ \\
\textbf{Isotropic}    & $-16.9(1)$ & $-17.7(1)$ & $-17.0(1)$ \\
\bottomrule
\end{tabular}
\begin{tablenotes}
\item \justifying \textit{Note}: The numbers in the parentheses denote the 1$\sigma$ errors in the last digit.
\end{tablenotes}
\caption{\justifying Summary of logarithmic evidence ($\ln \mathbb{Z}$) from \texttt{UltraNest} posterior sampling, comparing three EOS models (BSk22, MPA1, AP3) across distinct stellar configurations. Columns represent distinct EOS variants (each column corresponds to one EOS), while rows denote physical scenarios: dark matter admixed neutron stars with anisotropic baryonic matter (DANS), pure BM models incorporating pressure anisotropy, and isotropic BM benchmarks.}
\label{tab:logZ}
\end{threeparttable}
\end{table}

Bayesian evidence, denoted as $\mathbb{Z} = p(\boldsymbol{d}\,|\,\mathbb{M})$, quantifies the overall probability that a given model $\mathbb{M}$ can produce the observed data $\boldsymbol{d}$, marginalizing over the entire parameter space. A larger value of $\mathbb{Z}$ indicates stronger support for the model from the data. The Bayes' factor, defined as the ratio between the Bayesian evidence of two models,
\begin{equation}\label{eq:Bayes_factor}
    \mathcal{B}_{\mathrm{I}-\mathrm{II}}=\mathbb{Z}_{\mathrm{Model-I}}/\mathbb{Z}_{\mathrm{Model-II}},
\end{equation}
offers a direct quantitative measure between two models. For $\text{Model-I}$, Bayes' factors that are greater than 3.2 are typically interpreted as "substantially preferred", while those exceeding 10 are considered "strongly preferred."

To be specific, we performed Bayesian inference for two benchmark models: BM isotropic and BM anisotropic, alongside the DANS case. We summarized the resulting values, $\ln(\mathbb{Z})$ in Table~\ref{tab:logZ} and corresponding Bayes' factor between any two models in Figure~\ref{fig:bayesevidence_comp}, with their corresponding distributions of posteriors provided in the Appendix. This allows us to assess, for instance, whether DANS configurations are statistically favored/disfavored relative to pure baryonic NS under isotropic and anisotropic stress.

\begin{figure}[htbp]
    \centering
    \includegraphics[width=\columnwidth]{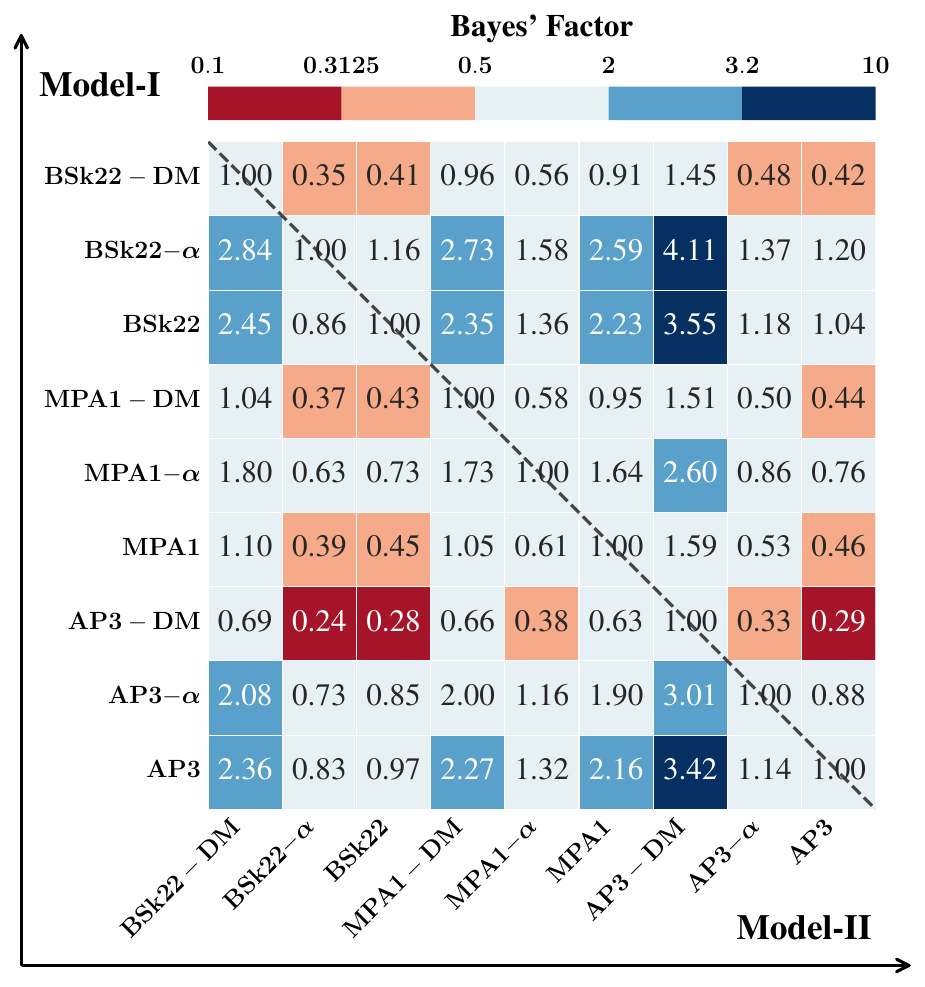}
    \caption{\justifying Heatmap illustrating pairwise model comparisons via Bayes' factors, where all values follow the definition in \eqref{eq:Bayes_factor}. Model labels on the vertical axis (left) denote the numerator (Model-I), and those on the horizontal axis (bottom) denote the denominator (Model-II) in the Bayes' factor ratio. A dashed line divides the matrix into upper and lower triangles, which exhibit reciprocal values ($\mathcal{B}_{\mathrm{I-II}} = 1/\mathcal{B}_{\mathrm{II-I}}$) as required by the definition of Bayes' factor. Each label commences with the EOS abbreviation (e.g., BSk22, MPA1, AP3), succeeded by suffixes indicating physical modifications. The '-DM' suffix indicates DANS, while '-$\alpha$' signifies the inclusion of pressure anisotropy. Labels without hyphens represent the pure isotropic benchmark cases. Deep blue (red) regions highlight that Model-I (-II) is significantly favored, with values exceeding 3.2 (or falling below $1/3.2=0.3125$).}
    \label{fig:bayesevidence_comp}
\end{figure}

Overall, DANS models consistently yield lower Bayesian evidence compared to their BM-only counterparts, primarily due to the additional degrees of freedom introduced by three or four extra parameters. However, the Bayes' factors for BM-only models never exceed 3.2 when compared to DANS models using either the BSk22 or MPA1 EOS, indicating that DANS models retain certain advantages. This is particularly evident in the low-mass region where the formation of DM halos can improve consistency with observations, as will be thoroughly analyzed in Sec.~\ref{sec:mac_prop_mr} through mass-radius relation diagnostics.

When examining fixed configurations across different EOS models, we find minimal variations in Bayesian evidence, suggesting current data cannot strongly discriminate between BM compositions. However, for fixed EOS with varying configurations, anisotropic models slightly outperform both isotropic and DANS models for BSk22 and MPA1, while for the softer AP3 EOS, isotropic models not only significantly favor over DANS models, but also slightly surpass anisotropic versions. This strong rejection of DM components in AP3 likely stems from its intrinsically soft EOS: when already marginally consistent with observations, additional DM softening drives the stellar configurations beyond observational limits. Conversely, for stiffer EOS like BSk22 and MPA1, DM-induced softening or halo structures provide beneficial adjustments. This suggests that DM admixing may be better accommodated in stiffer EOS frameworks.

\subsection{Close correlation between $R \chi$ and DM parameters}\label{sec:mac_prop_rx}
\begin{figure}
    \centering
    \includegraphics[width=\columnwidth]{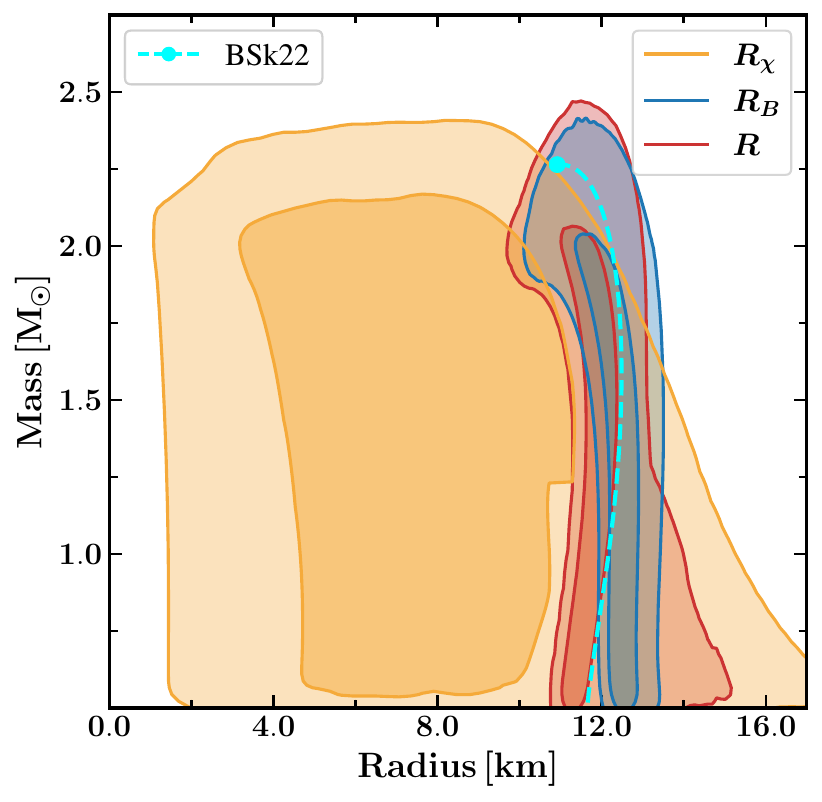}
    \caption{\justifying The posterior distributions of the characteristic radii $R_\chi$, $R_B$ and $R$ each versus $M$ for the BSk22-DM EOS, shown as a representative case. The shaded regions indicate the 68\% and 99\% credible regions, with yellow, blue, and red corresponding to $R_\chi$, $R_B$ and $R$, respectively. The dashed cyan line with a marker represents the $M$-$R$ relation for the corresponding isotropic solution.}
    \label{fig:R0R1R2M}
\end{figure}

\begin{figure}[h!]
    \centering
    \includegraphics[width=\columnwidth]{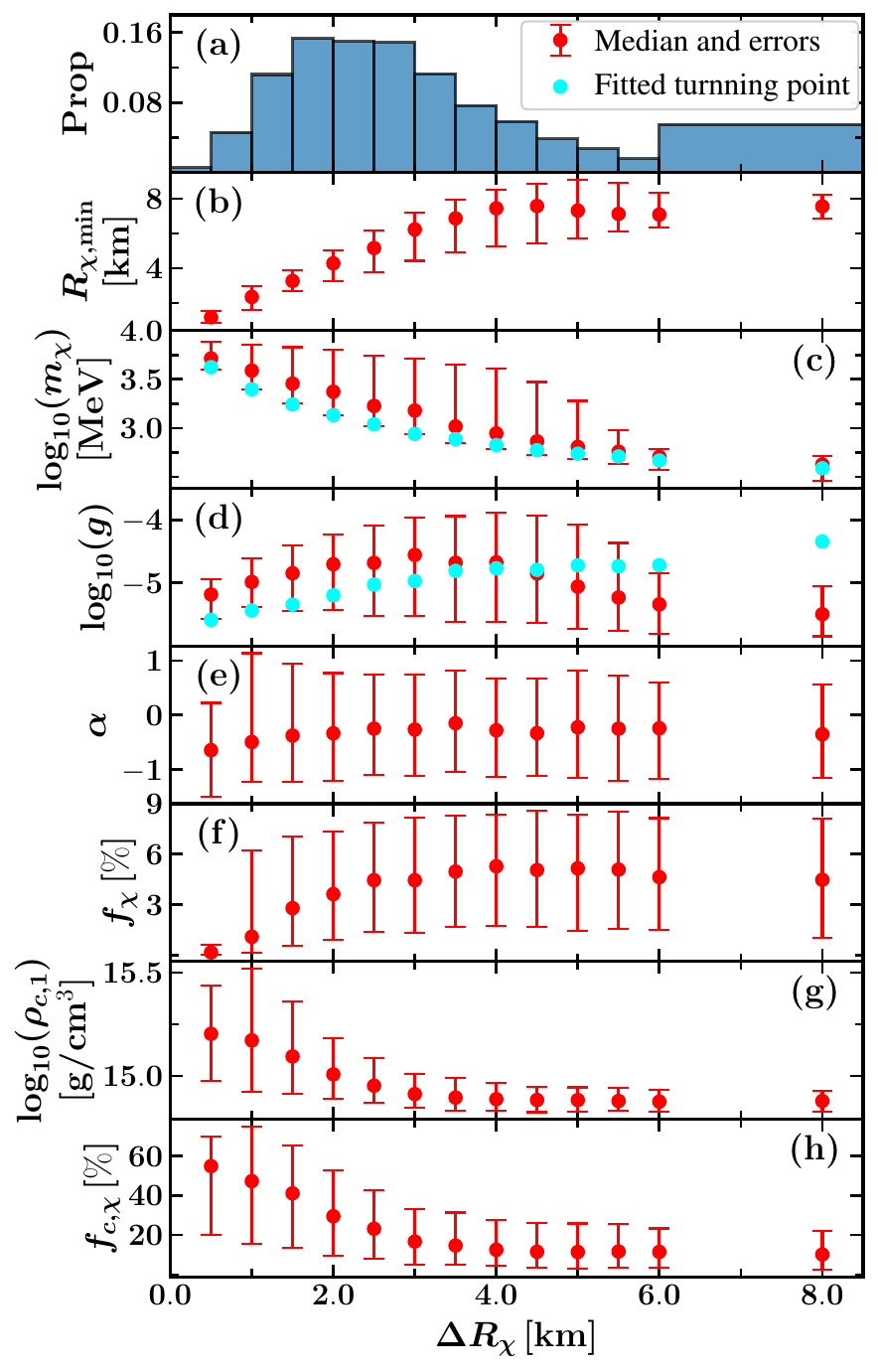}
    \caption{\justifying Parameter distributions of BSk22-DM across $\Delta R_\chi$ intervals. Panel (a) shows classification bins where box boundaries represent $\Delta R_\chi$ ranges ($\SI{8.5}{km}$ denotes $\infty$) and heights indicate subset proportions. Panels (b)--(h) display posterior distributions via red error bars (16\textsuperscript{th}/50\textsuperscript{th}/84\textsuperscript{th} percentiles), positioned at upper bin boundaries for finite intervals ($\Delta R_\chi < \SI{6}{km}$) or $\SI{8}{km}$ for the $[6,\infty)\,\si{km}$ case. Cyan points in (c)--(d) mark piecewise linear fits to the joint $\log_{10}(m_\chi)$--$\log_{10}(g)$ distributions.}
    \label{fig:R2_with_param}
\end{figure}

Although the DM radius, $R_\chi$, cannot yet be measured directly, its posterior distribution offers a window that is more tightly coupled to the underlying particle properties than global stellar observables. In this section, we analyze the $R_\chi$ posteriors and their systematic correlations with $m_\chi$ and $g$. This mapping will be pivotal once future data allow $R_\chi$ to be inferred, whether through high-precision pulse profile modeling or characteristic gravitational-wave signatures, because it will convert those measurements into refined limits on the admissible DM parameter space, thereby opening a new avenue for testing DM models in compact stars.

Figure~\ref{fig:R0R1R2M} shows the joint posteriors for the DM radius $R_\chi$, the baryonic radius $R_B$ and the total stellar radius $R$ each versus the gravitational mass $M$ in our DANS models. The sampled points reveal a clear separation between the radii: while $R_B$ and $R$ are tightly clustered around $\sim\!\SI{12}{km}$, $R_\chi$ is both systematically smaller and broader. These trends indicate that the present data and priors favor configurations in which a compact DM core coexists with a baryonic envelope whose radius is well determined, yet they also underscore the large residual uncertainty in the internal DM distribution. 

Part of this preference arises from an implicit selection bias introduced by our choice of the empirical radius definition from \cite{collier2022tidal}: configurations that develop extended DM haloes redistribute mass toward larger radii, pushing $R$ beyond the range we allow and thereby being discarded during the sampling. Within this framework the sampled $M$–$R_B$ pairs trace the mass–radius relation of the underlying BSk22 EOS almost unchanged, showing that DM admixtures with approximately 10\% of $f_\chi$ leave the $R_B$ largely unaffected. If we were to relax the empirical definition and instead equate the stellar radius directly with $R_B$, the favored parameter space would continue to satisfy the observational constraints, although it would naturally broaden to admit halo-dominated solutions.

The dominance of DM–core solutions also shapes the joint posterior in the $\log_{10}(m_\chi)$–$\log_{10}(g)$ plane (Figure~\ref{fig:param_all}). Analytical estimates for an isotropic DANS model (Figure~3 of \cite{miao2022dark}) show that this plane is naturally partitioned into three regimes: (\romannumeral1) parameter pairs that generate a self-gravitating DM core, (\romannumeral2) pairs that lead to an extended DM halo, and (\romannumeral3) combinations where the overall maximum mass falls below observational limits. Our samples occupy exclusively the first of these regions, producing the broken-line boundary evident in Figure~\ref{fig:param_all}.

This feature arises from the same selection bias discussed above: configurations that would form a halo shift baryonic mass outward, increasing the stellar radius beyond the empirically allowed range and thus being disfavored in the sampling. Consequently, the current posterior constrains $m_\chi$ and $g$ to values compatible with compact cores only. Future analyses aimed at testing halo-dominated scenarios should therefore relax the empirical radius criterion and instead interpret forthcoming x-ray or gravitational-wave data within a framework that admits both core and halo configurations.

These results demonstrate that neither empirical radius definitions nor global $M$-$R$-$\Lambda$ observations alone can effectively constrain the DM parameters $m_\chi$ and $g$ to narrow ranges. Given the wide distribution of $R_\chi$ and its potential impact on DANS cooling processes, we leverage Bayesian analysis to systematically study $R_\chi$ distributions. This approach will facilitate future constraints on $m_\chi$ and $g$ when probes sensitive to $R_\chi$ become available.

To quantify how a given DM parameter set can reshape the internal structure of a star, we define   
$
\Delta R_\chi \;=\; R_{\chi,\max}-R_{\chi,\min},
$
the total range of DM radii obtained as $\rho_c$ is scanned over all values that yield equilibrium configurations compatible with our priors (Sec. \ref{sec:prior}). A single radius evaluated at a canonical mass would miss much of this structural latitude, whereas $\Delta R_\chi$ compresses the entire response of the DM size into one diagnostic number. Small $\Delta R_\chi$ signals a tightly constrained core, while large values flag parameter choices that permit halolike expansions. Because empirical EOS constraints are trustworthy only for NSs heavier than $\SI{0.5}{M_\odot}$, the extrema are taken from models in this mass range. The examples below adopt the BSk22-DM EOS, which typifies the behaviour seen across our full ensemble.

We divide the $\Delta R_\chi$ axis into thirteen mutually exclusive bins: twelve consecutive intervals of width $\SI{0.5}{km}$ covering $0$–$\SI{6}{km}$, plus a final open bin for $\Delta R_\chi \ge \SI{6}{km}$.  
Figure~\ref{fig:R2_with_param} summarizes the resulting subsets.  
Panel~(a) shows the fractional occupancy $N_i/N$ of each bin, while panels~(b)–(h) plot the posterior distributions of key model parameters and some physical quantities within those bins.  
Error bars are placed at the upper edge of each $\SI{0.5}{km}$ interval and at $\SI{8}{km}$ for the wide tail bin, whose mean value is $\langle\Delta R_\chi\rangle\simeq\SI{7.8}{km}$.%

This visualization reveals both the constrained parameter ranges within individual $\Delta R_\chi$ bins and their progression across increasing $\Delta R_\chi$ magnitudes. Global observables such as the gravitational mass $M$, stellar radius $R$, and tidal deformability $\Lambda$ are omitted, because their posteriors change only marginally with $\Delta R_\chi$ and thus provide little insight into this otherwise hidden DM property.

Panels (a) and (b) collectively characterize the distribution and evolution of $R_\chi$ with $\Delta R_\chi$ in the DANS model, revealing a critical transition at $\Delta R_\chi \simeq \SI{4}{km}$. Panel (a) shows $\Delta R_\chi$ values predominantly concentrate in the 1--4\, km range, with only 5\% of samples exceeding $\SI{6}{km}$ (a conservative cutoff imposed to mitigate sampling noise). Panel (b) further demonstrates that below $\Delta R_\chi \!\lesssim\! \SI{4}{km}$, the $R_{\chi,\min}$ increases steadily, reaching a central value of $\SI{8}{km}$ at the transition point. Above $\SI{4}{km}$, however, $R_{\chi,\min}$ plateaus at this $\SI{8}{km}$ value. Significantly, this transition coincides with $R_{\chi,\max} = \Delta R_\chi + R_{\chi,\min}$ reaching $\SI{12}{km}$, a scale matching the average radius from three NICER observations. This $\SI{12}{km}$ threshold thus marks the transition from compact DM core dominance to diffuse halo formation in DANS.

Panels (c) and (d) illustrate how $m_\chi$ and $g$ evolve with $\Delta R_\chi$. In panel (c), $m_\chi$ exhibits a monotonic decreasing trend as $\Delta R_\chi$ increases. Panel (d), however, reveals $g$ follows a biphasic behavior analogous to $R_{\chi,\min}$: below $\Delta R_\chi \!\lesssim\! \SI{4}{km}$, $g$ increases with $\Delta R_\chi$, yet above this threshold, it reverses to decrease with further $\Delta R_\chi$ growth. Since $m_\chi$ and $g$ directly govern the DM EOS, these trends imply distinct EOS evolutions across the transition: for $\Delta R_\chi \!\lesssim\! \SI{4}{km}$, the overall DM EOS stiffens, corresponding to concurrent increases in both $R_{\chi,\min}$ and $R_{\chi,\max}$. For $\Delta R_\chi \!\gtrsim\! \SI{4}{km}$, however, the EOS softens in the high-density core while stiffening in the low-density outer regions, consistent with $R_{\chi,\min}$ plateauing even as $R_{\chi,\max}$ continues to expand.

The two–segment, "broken-line" shape seen earlier in Figure \ref{fig:param_all} persists in the joint posteriors for the $m_\chi$ and $g$.  
To locate the knee of that bend quantitatively, we applied a single-break least-squares fit: for each $\Delta R_\chi$ subset, we fitted two straight lines with a shared breakpoint to the $m_\chi$-$g$ data, minimizing the sum of squared residuals defined as $\Delta(\log_{10}(m_\chi))$ at fixed $g$ values. The resulting breakpoints are plotted as cyan circles in panels~(c) and (d).   For subsets with $\Delta R_\chi\le\SI{6}{km}$ the fitted breakpoints fall almost exactly on the medians obtained in Figure \ref{fig:param_all}, confirming the internal consistency of the two analyses.  In the widest-span class ($\Delta R_\chi>\SI{6}{km}$), however, a departure from the earlier median trend that disrupts the linear two-segment fitting model. This manifests as a vertically elongated covariance ridge in the $m_\chi$–$g$ plane, signaling that halo-dominated configurations impose a distinctly different correlation structure on the DM parameters compared to core-dominated systems.

To investigate how the microscopic DM parameters $m_\chi$ and $g$ influence BM properties, we next plot the posterior of $\alpha$, $f_\chi$, $\log_{10}(\rho_{c,1})$ (chosen as a representative for the $\rho_{c,i}$), and $f_{c,\chi}$ in panels~(e)-(h).
In the compact-core regime ($\Delta R_\chi<\SI{4}{km}$), all four quantities vary almost linearly with $\Delta R_\chi$, mirroring the steady expansion of the core captured by $R_{\chi,\min}$.  Once $\Delta R_\chi$ exceeds $\SI{4}{km}$, the curves flatten, indicating structural saturation despite continued halo expansion. This contrasts sharply with the evolving microscopic parameters $m_\chi$ and $g$, revealing a critical decoupling: in the halo-formation regime, particle-level EOS adjustments no longer influence large-scale stellar geometry. Instead, within the low $m_\chi$, low $g$ domain, variations in $(m_\chi,g)$ affect only DM-specific measures like $R_\chi$ or $\Delta R_\chi$. This EOS decoupling from baryonic structure implies that distinct combinations of light, weakly interacting DM particles can yield nearly identical observable NS properties while imprinting large differences in the hidden DM halo.

Zooming into the parameter trends of panels (e)–(h) for $\Delta R_\chi < \SI{4}{km}$, we first focus on the smallest spans ($\Delta R_\chi < \SI{2}{km}$). Here, the soft DM EOS drives dense core concentration, reflected in elevated $f_{c,\chi}$ values. This configuration requires both a reduced $f_\chi$ and an increased $\rho_{c,1}$ to maintain stellar stability and suitable $M$ and $R$, which explains the origin of the higher-density tails (68\%--99\% credible intervals) in Figure~\ref{fig:param_all}. As $\Delta R_\chi$ grows larger, the DM EOS stiffens, producing a broader distribution of $f_\chi$ values. However, the concurrent slight increase in $\alpha$ suggests that the softening influence from the amount of DM mass partially compensates for the DM EOS hardening, maintaining a balance between these competing effects.

In summary, present multimessenger constraints on the $M$, $R$, and $\Lambda$ leave a degeneracy between  $m_\chi$ and $g$ in DANSs. By contrast, $R_\chi$ or $\Delta R_\chi$ responds directly and distinctly to each of these microscopic parameters. Future observations capable of inferring $R_\chi$ therefore hold the key to disentangling $m_\chi$ from $g$ and placing independent constraints on the DM sector.

\subsection{Macroscopic properties of DANS}\label{sec:mac_prop_mr}
\begin{figure}
    \centering
    \includegraphics[width=\columnwidth]{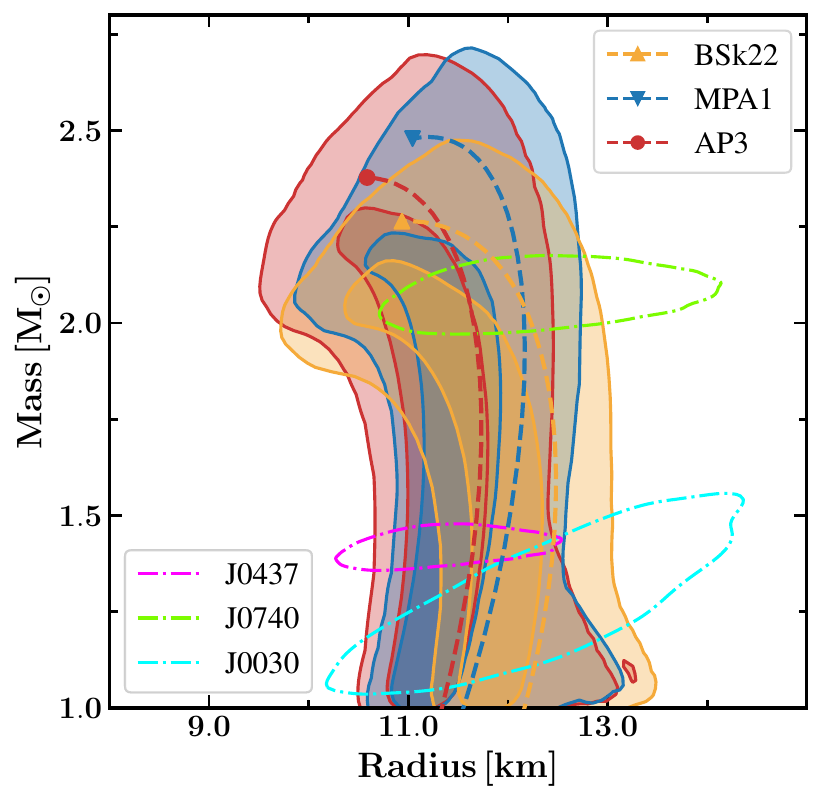}
    \caption{\justifying The $M$-$R$ posteriors for different EOS, with 68\% and 99\% credible regions indicated by shaded contours. Dot-dashed curves mark 68\%  confidence constraints from NICER for selected pulsars (J0030, J0740, J0437), while dashed lines trace the BM-isotropic solutions. The BM components BSk22, MPA1, and AP3 are color-coded for clarity.}
    \label{fig:MR_DANS}
\end{figure}
\begin{figure}
    \centering
    \includegraphics[width=\columnwidth]{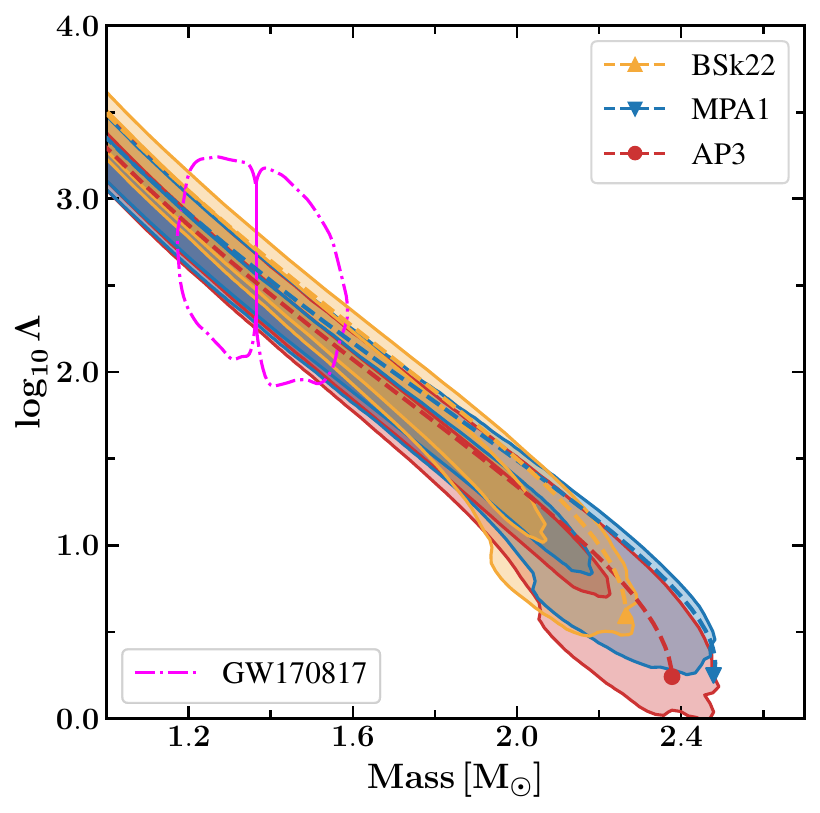}
    \caption{\justifying The posterior distributions for the $M$-$\Lambda$ relation. The dash-dotted magenta curve indicates the 68\% confidence constraint from GW170817. The shaded contour and dashed line conventions follow those in Figure \ref{fig:MR_DANS}, and the color scheme remains consistent.}
    \label{fig:ML}
\end{figure}

To more intuitively visualize how DM modifies different BM components, we now examine the posteriors of macroscopic observables, specifically the mass-radius ($M$-$R$) and mass-tidal deformability ($M$-$\Lambda$) relation. Figs.~\ref{fig:MR_DANS} and \ref{fig:ML} show the $M$-$R$ and $M$-$\Lambda$ posterior distributions. The contours are overlaid with observational constraints from x-ray for millisecond pulsars (J0030, J0740, J0437), and from the GW signal GW170817.

Figure~\ref{fig:MR_DANS} shows the marginalized posteriors in the $M$–$R$ plane for DANS constructed with three representative nuclear EOSs. In every case, the 68\% credible regions of DANS contours lie to the left of their baryonic counterparts, indicating smaller radii at the same mass. The shift has a simple physical origin: DM contributes to the total mass but supplies almost no pressure, so gravity is stronger relative to the pressure that can be provided by BM alone. However, the 68\%--99\% credible regions spread even exceeded the BM curve from left to right, particularly below $\SI{1.5}{M_\odot}$. This arching feature indicates DM halo formation, and in this way, while satisfying the radius constraints from J0740 and J0437 for stars above $\SI{1.25}{M_\odot}$, DANS models simultaneously allow larger radii in the low-mass regime, bringing the $M$-$R$ curves closer to the central values of J0030--—a feature unattainable in BM-only models. This demonstrates DANS' unique ability to better reconcile theoretical predictions with observational data.

Figure~\ref{fig:ML} shows the posterior density in the $M$–$\Lambda$ plane for DANSs.  
At any given $M$ the contours shift toward lower $\Lambda$ relative to BM-only models, echoing the increase in compactness already seen in the $M$–$R$ diagram.  
Because $\Lambda$ scales as $C^{-5}$ with compactness $C=M/R$, even a modest DM-induced reduction in $R$ drives a pronounced suppression of $\Lambda$. DM therefore renders the star both smaller and less deformable than its baryonic counterpart.

Multimessenger data, especially the $\Lambda$ limit from GW170817, make the DM effect on EOS important. A small DM component shrinks the star and lowers $\Lambda$, rescuing EOSs that would otherwise overshoot the bound, but it can also push compliant EOSs below it. DM thus shifts every BM EOS toward smaller $R$ and $\Lambda$; whether this improves or worsens agreement with observations depends on the EOS's original location relative to the measured band.

Current $M$–$R$ and $M$–$\Lambda$ measurements allow NS configurations with $\mathcal{O}(1\%)$ DM mass fractions. At the same time, the DANS sequences depart systematically from BM-only relations in both the $M$–$R$ and $M$–$\Lambda$ planes, offering a clear observational handle on the DM sector. Forthcoming high-precision data from NICER, LIGO/Virgo/KAGRA, and next-generation detectors such as the Einstein Telescope \cite{Maggiore_2020} and Cosmic Explorer \cite{reitze2019cosmicexploreruscontribution} should therefore be able to confirm, or rule out, these DM-admixed configurations.%

\subsection{EOS and phase transitions in DANS}
\begin{figure}[htbp]
    \centering
    \includegraphics[width=\columnwidth]{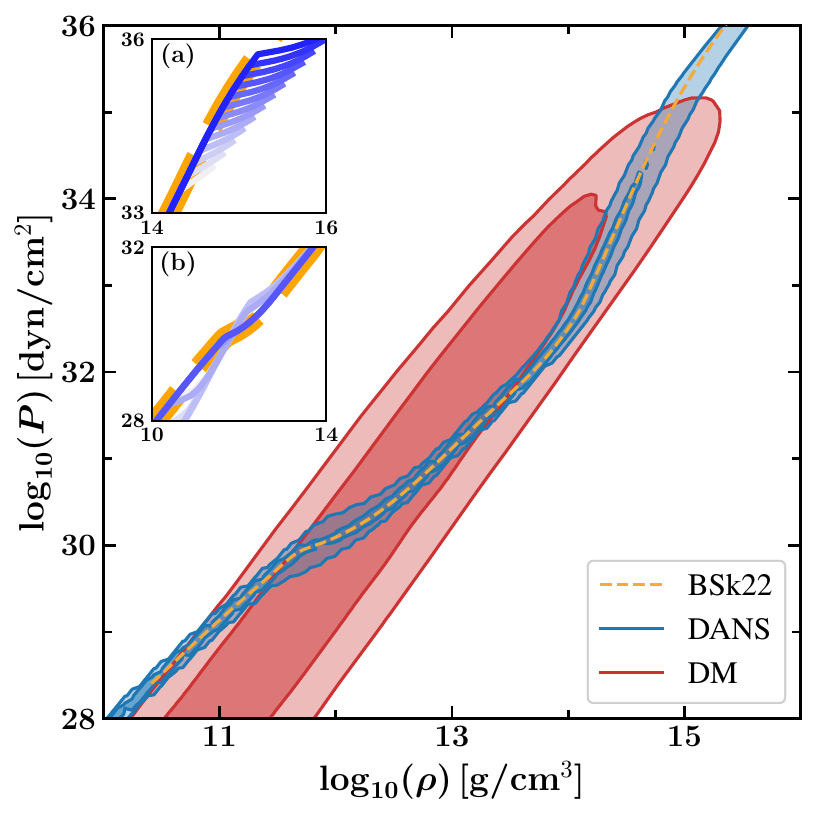}
    \caption{\justifying The $P$-$\rho$ relations derived from the posterior distributions for the BSk22-DM EOS. The main panel shows 68\% and 95\% credible regions with blue contours representing the overall DANS $P_T$–$\rho_T$ relation, while red contours maintaining 68\% and 99\% credible levels correspond to the DM $P_\chi$–$\rho_\chi$ relation. Insets (a) and (b) zoom into the core and crust regions, respectively, highlighting 13 representative EOS realizations extracted from the posterior, color-coded by increasing central density $\rho_c$. These realizations illustrate the diversity in the EOSs within the DANS distribution. For comparison, the dashed orange curve denotes the BSk22-isotropic EOS.}
    \label{fig:EOS}
\end{figure}

\begin{figure}[htbp]
    \centering
    \includegraphics[width=\columnwidth]{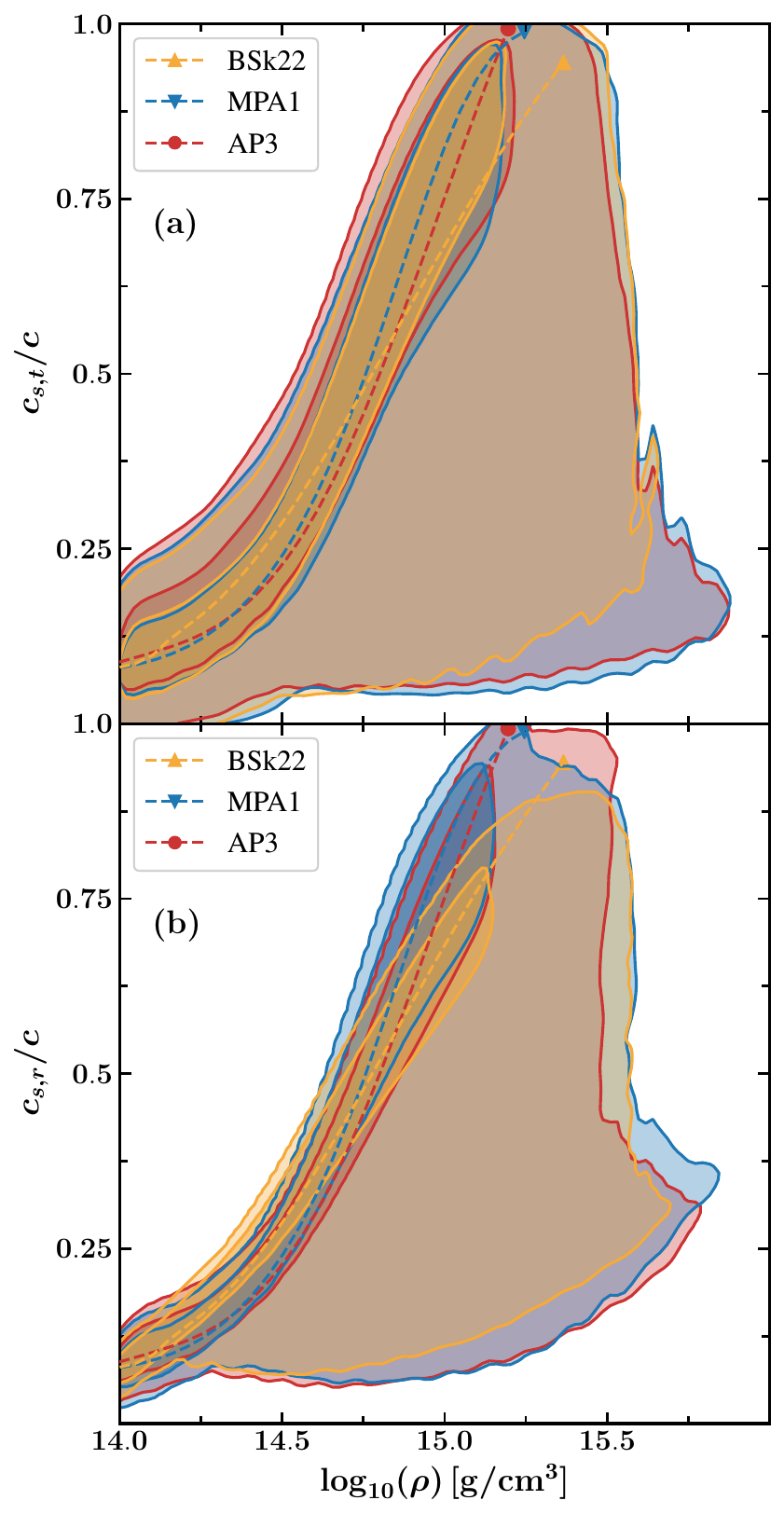}
    \caption{\justifying The posterior distributions for tangential speed of sound $c_{s,t}$-$\rho$, where $c_{s,t}$ has been normalized with the speed of light $c$ and we take the logarithm of $\rho$. The shaded contours follow the same convention as in Figure \ref{fig:MR_DANS}, while the dash-dotted lines represent the radial speed of sound $c_{s,r}$ for BM-isotropic cases.}
    \label{fig:Cst}
\end{figure}

DM modifies a NS's microphysics, especially its EOS. From the posterior EOS ensemble, we track the pressure-density ($P$-$\rho$) curve and speed of sound $c_s(\rho)$, linking microphysical changes to macroscopic shifts. The stiffening or softening of EOS, and even the possible phase-transition signals inside DANS from DM-BM interplay, can now be tested with multimessenger observations.%

Figure~\ref{fig:EOS} compares the EOS inferred for the BSk22 with and without DM. In the main panel the red contour band shows the DM-only relation $P_\chi(\rho_\chi)$, while the blue band gives the total pressure $P_T(\rho_T)$ of DANS; the dashed yellow line is the isotropic BM-only EOS of BSk22 for reference. Insets~(a) and (b) magnify the core and crust regions respectively, and overlay 13 representative EOS draws from the posterior, colored from light to dark according to increasing central density $\rho_c$. These samples illustrate how DM softens the core at high~$\rho_c$, yet leaves the low-density crust unchanged in most cases, linking the microphysical deviations to the macroscopic $R$ and $\Lambda$ shifts discussed earlier.

In our two-fluid fermionic DM model, the $P-\rho$ relation of the DM component depends on $g$ and $m_\chi$, producing a relatively broad, bandlike structure in $\log_{10}P$-$\log_{10}\rho$ space. However, the total DANS EOS remains narrowly clustered near the original BSk22 relation, reflecting the prior constraint $f_\chi < 10\%$ which limits DM's total mass contribution. While $P_\chi(\rho)$ exceeds $P_B(\rho)$ at specific densities (e.g., $\SI[parse-numbers=false]{10^{13}}{g/cm^3}$), $P_T$ in DANS remains comparable to $P_B$. This occurs because $\rho_B$ typically dominates over $\rho_\chi$ by an order of magnitude in most regimes, leading to the approximate relation:
\begin{equation}\label{eq:P_T_P_B_P_chi}
    P_T(\rho_B + \rho_\chi) = P_B(\rho_B) + P_\chi(\rho_\chi) \sim P_B(\rho_B + \rho_\chi)
\end{equation}
Nevertheless, deviations from this approximation emerge in two specific regimes that are,
\begin{itemize}
    \item\textbf{Core region} ($\rho \gtrsim \SI[parse-numbers=false]{10^{14}}{g/cm^3}$):
    In Figure~\ref{fig:EOS} (a), all 13 EOS curves exhibit downward folding in the core region relative to the isotropic BSk22 reference (yellow dashed line), indicating DM-induced softening. While $\rho_\chi$ becomes non-negligible at this region, the DM EOS becomes softer than the BM EOS ($P_\chi(\rho) < P_B(\rho)$). From the BM's perspective, this softness implies that the DM pressure contribution cannot compensate for the BM pressure, resulting in $P_T$ satisfying:
    \begin{equation}
        P_T(\rho_B + \rho_\chi) = P_B(\rho_B) + P_\chi(\rho_\chi) < P_B(\rho_B+\rho_\chi)
    \end{equation}
    \item\textbf{Crust region} ($\rho \sim \SI[parse-numbers=false]{10^{12}}{g/cm^3}$):
    In Figure~\ref{fig:EOS} (b), the $P$-$\rho$ relation exhibits pressure enhancement in high-density regions and suppression in low-density regions, reflecting the dominant influence of DM. However, only four low-$\rho_c$ realizations deviate from the BM baseline in this representative case (versus nine overlapping curves). This implies that significant DM impact in the crust is rare, requiring both a low value of $\rho_c$ and specific DM parameter combinations to accumulate appreciable DM in the outer layers.
\end{itemize} 

The Figure~\ref{fig:Cst} illustrates the posterior distributions of the tangential and radial speed of sound $c_{s,t}$ and $c_{s,r}$ (normalized to the speed of light $c$) versus density $\log_{10}{\rho}$ for DANS. The distributions are shown for three representative BM, while the dash-dotted lines represent the corresponding $c_{s,r}$ for BM-only configurations.

The posteriors of $c_{s,t}$ and $c_{s,r}$ increase with $\rho$ in the $68\%$ confidence regions, but a pronounced broadening of the posterior is evident at higher densities. Notably, this extended high-density structure is absent in the corresponding BM-only case (see Figure~\ref{fig:Cst_alpha} in the Appendix), confirming that the additional softening feature in Figure~\ref{fig:EOS} is a potential first-order phase transition induced by the presence of DM. Although the microscopic origin of this phase transition remains uncertain, our analysis demonstrates that lower values of the $f_\chi$, $m_\chi$, and higher values of $g$ tend to reduce the abruptness of $c_s$ transitions in core regions. This behavior is physically understandable: when either (i) the overall DM fraction decreases, or (ii) the DM EOS stiffens (leading to reduced $\rho_\chi$ in the core region), the DM's influence on the core EOS will become attenuated. 

Another intriguing feature emerges when comparing the $c_{s,t}$ and $c_{s,r}$ profiles. While their overall morphology remains consistent, subtle differences appear in their 68\% credible regions. For all three BM components in Figure~\ref{fig:Cst} (a), the $c_{s,t}$ credible regions cluster together with nearly identical contours. In contrast, Figure~\ref{fig:Cst} (b) reveals that the $c_{s,r}$ regions instead align distinctly along their respective isotropic BM-only $c_s$ curves. This suggests that $c_{s,t}$ exhibits weaker dependence on BM composition compared to $c_{s,r}$. As evident from \eqref{eq:anisoBL}, this behavior likely reflects how the anisotropic pressure parameter $\alpha$ mitigates BM-specific effects, thereby driving the $P_t$-$\rho$ relation toward a universal trend across different EOS.

\section{Conclusion}\label{sec:conclusion}
In this study, we developed a two-fluid model of DANS where DM and BM interact solely through gravitational coupling. Here, we assumed BM to be anisotropic and DM to be isotropic. To capture the uncertainties in BM properties and the complex stellar interior environment, we incorporated three different types of BM EOS and introduced anisotropic pressure effects. Given the current lack of DM considerations in x-ray and GW analyses, we defined the stellar radius as enclosing 99.99\% of the total mass. Through Bayesian analysis, combining our framework with multimessenger observations from NICER mass-radius measurements and tidal deformability from GW170817, several key findings emerged.

Although current data place almost no upper limit on $f_\chi$, examining the joint posterior of $m_\chi$ and $g$ initially produced a distinctive broken-line pattern.  Further inspection, however, shows that this feature is driven mainly by our empirical radius definition, which systematically disfavors configurations with extended DM halos and instead selects for compact cores. By contrast, the anisotropy parameters in the BM sector are much better constrained since BM dominates the mass contribution and directly determines the overall $M$-$R$ relation.  Even with 3 extra degrees of freedom from DM, Bayes factors do not favor BM-only models over DANSs. Finally, whether DM admixture actually improves agreement depends on the BM EOS: soft EOS candidates like AP3 disfavor DM inclusion, whereas stiffer EOSs (BSk22 and MPA1) can accommodate a $\mathcal{O}(1\%)$ DM component that shifts their stiff BM EOS toward observational limits.

Because our empirical radius definition disfavors extended DM haloes, $R_\chi$ always remains below $R_B$, but it can span a wide range of smaller values. To prepare for future indirect measurements of $R_\chi$ that may resolve the current $(m_\chi,g)$ degeneracy, we categorize models according to $\Delta R_\chi$.
For $\Delta R_\chi < \SI{4}{km}$, we find that $R_{\chi,\min}$ correlates nearly linearly with $R_\chi$, $m_\chi$, and $g$.  Once $\Delta R_\chi$ exceeds $\SI{4}{km}$, the behavior changes: $m_\chi$ maintains its decreasing trend, whereas $g$ exhibits a reversed trend and $R_\chi$ stabilizes. Although the selection bias remains active in this halo-dominated regime, the tight correlations established for $R_\chi < \SI{8}{km}$ suggest that even partial information about $R_\chi$ could significantly sharpen constraints on the DM parameters.

The primary astrophysical impact of DM in DANS involves EOS softening, leading to systematically smaller $R$, $M$, and $\Lambda$. Our models predict DM halo formation only when the $M$ below $\SI{1.5}{M_\odot}$, as stronger gravitational compression dominates in more massive stars. Microphysical analysis through the $P$-$\rho$ relation reveals that above $\SI[parse-numbers=false]{10^{14}}{g/cm^3}$, DM softens the EOS and induces first-order phase transitions. This mechanism eliminates the superluminal speed of sound that would otherwise occur in high-density BM regions, maintaining physical consistency through post-transition speed of sound reduction.

Our study provides a foundational advance in modeling DANSs. Because DM interacts only via gravity and does not couple to electromagnetic fields, GW tidal signatures depend exclusively on stellar compactness, not composition. By including pressure anisotropy alongside a self-interacting DM component, our framework captures the internal structure and dynamics of compact stars more realistically.  This enhancement improves consistency with existing $R$ and $\Lambda$ limits, sharpens predictions for multimessenger signals, and extends the reach of compact-object physics into the DM sector.

Nonetheless, other configurations, such as isotropic BM with anisotropic DM, or both components being anisotropic, are also viable and merit investigation in future work. As highlighted in \cite{Hippert:2022snq}, the stability of NSs is closely tied to the density distributions of both BM and DM. Future detectors like Advanced LIGO, LISA \cite{LISA_2017pwj}, the Einstein Telescope \cite{maggiore2020science}, and Cosmic Explorer \cite{reitze2019cosmic} are therefore ideally suited to reveal even subdominant DM cores or halos in NSs. Observations of high-mass pulsars, such as PSR~J0952-0607 \cite{Romani:2022jhd}, present a challenge for traditional isotropic models, which often fail to account for such extreme masses. In contrast, anisotropic models offer a more plausible framework, allowing for stable NS configurations at higher masses. Moreover, anisotropy influences the stellar deformation and tidal response, thereby enhancing the accuracy of gravitational waveform predictions. In this way, our work deepens understanding of extreme matter and paves the way for future multi-messenger studies aimed at uncovering exotic particles inside NSs.

\section*{Acknowledgement}
The authors sincerely thank the anonymous reviewer for their valuable comments and suggestions, which have significantly improved this manuscript. P.M would like to thank BITS Pilani, K K Birla Goa campus for the fellowship support; Saumyen Kundu for helping with how to run the Bayesian code and use of HRI-HPC for this work; Sayantan Ghosh for discussion about the phase transitions in NS; Subhadip Sau and Apratim Ganguly for related discussion on different DM models which have an impact on NS's two-fluid approach. X.Z thanks Jing-Yao Li for clarifying \texttt{emcee}/\texttt{UltraNest} differences and patience during occasional technical crises; Yong-Xiang Cui for multi-process programming guidance, MCMC fundamentals, and invaluable workstation optimization techniques; and Wei-Cheng Long for elegant Linux administration techniques and inspiring optimism.

\section*{DATA AVAILABILITY}
The data that support the findings of this article are openly available \cite{liu_2025_17082249}

\section*{Appendix: Posterior Results of BM Neutron Stars with and without Pressure Anisotropy}\label{append:iso&aniso}
For reference, we present the posterior distributions for BM-anisotropic and BM-isotropic configurations to illustrate how pressure anisotropy modifies NS properties. These results serve as a baseline for understanding the additional effects introduced by fermionic DM in the main analysis.

\begin{figure}[H]
    \centering
    \includegraphics[width=\columnwidth]{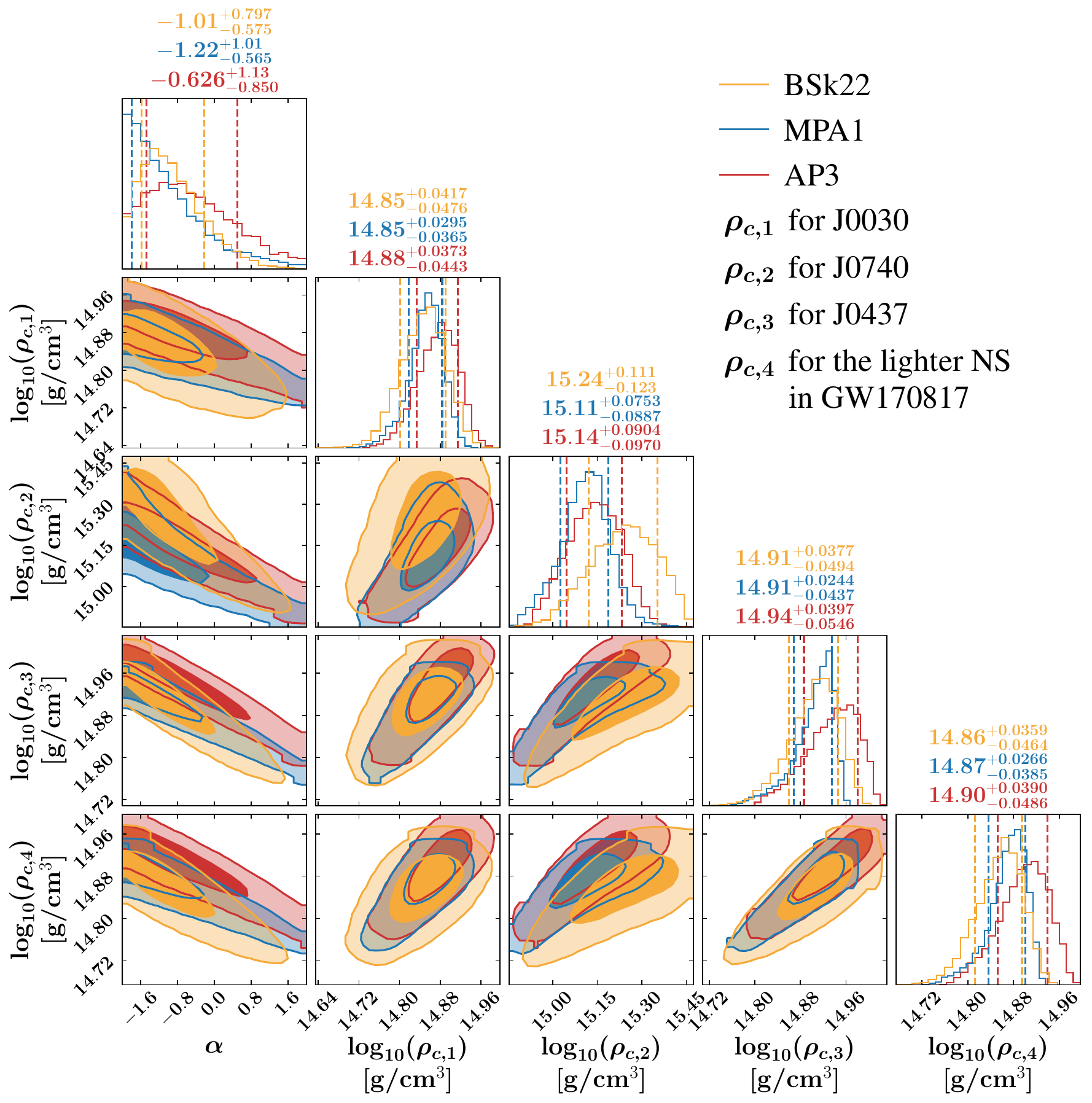}
    \caption{\justifying Posterior distributions for $\alpha$ and the logarithm of central densities for BM-anisotropic models. The shaded contours, dashed lines, and numerical annotations follow the same conventions as described for Figure~\ref{fig:param_all}}
    \label{fig:param_alpha_all}
\end{figure}

Figure \ref{fig:param_alpha_all} shows results for the BM-anisotropic model derived from the same observational constraints, where only the $\alpha$ parameter modifies the EOS. Compared to DM-containing models \ref{fig:param_all}, we observe modest shifts toward lower values in $\alpha$ and more concentrated distributions but discretized peaks for different BM EOS for the central density.

\begin{figure}[H]
    \centering
    \includegraphics[width=\columnwidth]{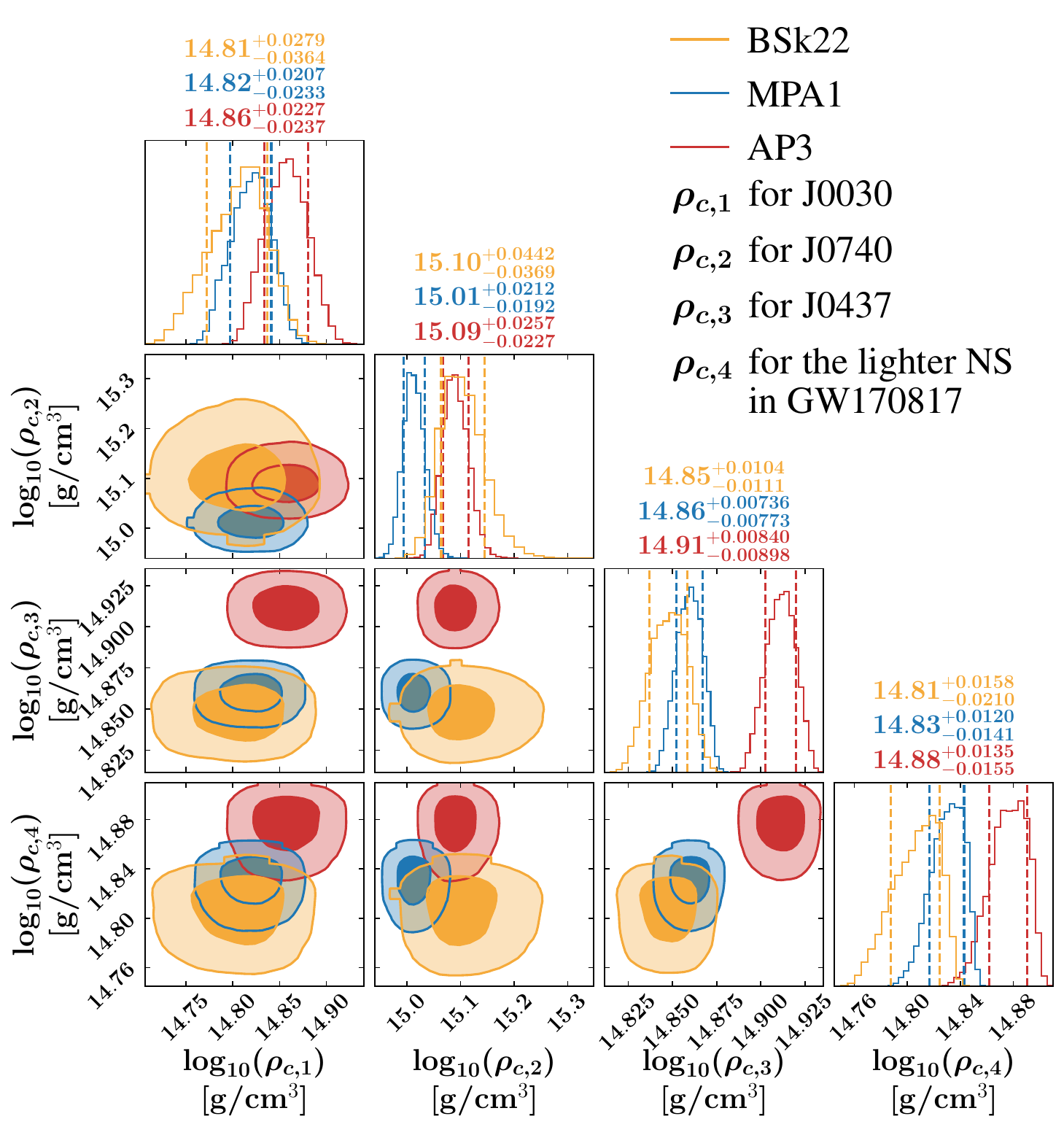}
    \caption{\justifying Posterior distributions for the logarithm of central densities for BM-isotropic models. The shaded contours, dashed lines, and numerical annotations follow the same conventions as described for Figure~\ref{fig:param_all}}
    \label{fig:param_EOS}
\end{figure}

Figure \ref{fig:param_EOS} shows results for the BM-isotropic model where central densities show tighter clustering for each EOS, but are clearly discrete for different EOS. All of their median values are smaller compared to models with DM of anisotropy.

\begin{figure}[htbp]
    \centering
    \includegraphics[width=\columnwidth]{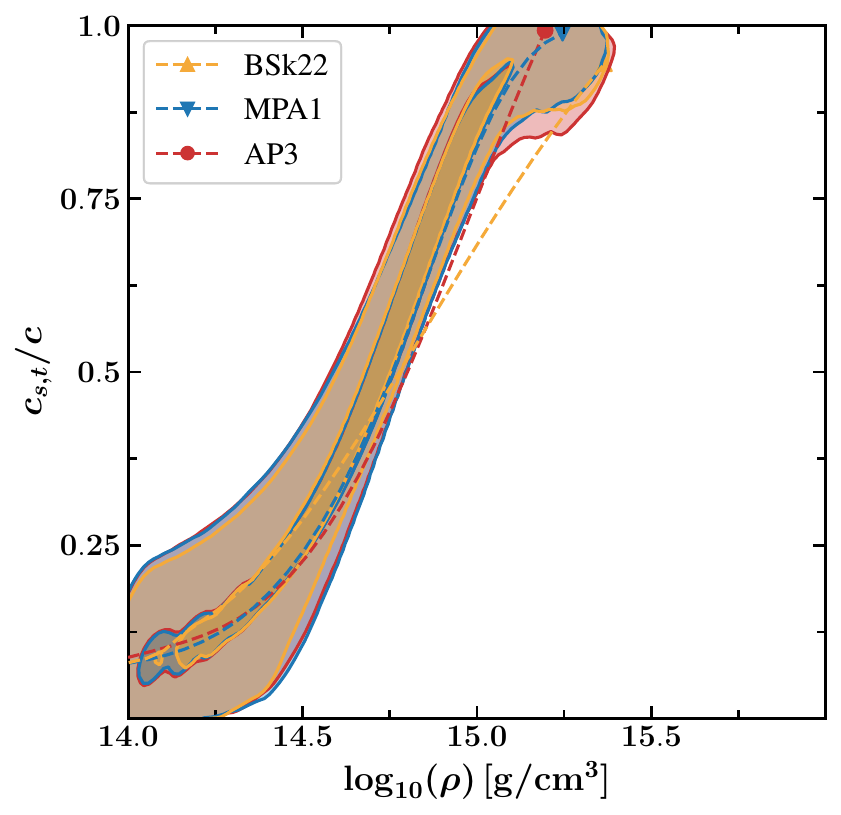}
    \caption{\justifying Posterior distributions of $c_{s,t}$-$\rho$ for anisotropic-BM models, where $c_{s,t}$ has been normalized with the speed of light $c$ and we take the logarithm of $\rho$. The shaded contours follow the same convention as in Figure \ref{fig:MR_DANS}}
    \label{fig:Cst_alpha}
\end{figure}

In Figure \ref{fig:Cst_alpha}, we also display the posterior of the $c_{s,t}$ for BM-anisotropic models. Without phase transition brought from DM, the posteriors are more concentrated around the $c_{s,t}$ curves of BM, compared to Figure \ref{fig:Cst}. It should be noted that although the median values of anisotropy for different BM EOS are quite different, their $c_{s,t}$ distributions are very similar to each other, even when anisotropy directly affects the value of tangential pressure $P_t$.


\FloatBarrier
\bibliography{citations}
\bibliographystyle{unsrt}
\end{document}